\documentclass{jpp}
\usepackage{graphicx}
\usepackage{epstopdf, epsfig}
\usepackage{hyperref}
\usepackage{amsmath,amssymb}
\usepackage{xspace}
\usepackage{xcolor}
\usepackage{multirow}

\newcommand{\dd}{\text{d}}
\newcommand{\red}{\textcolor{black}}
\newcommand{\blue}{\textcolor{black}}
\newcommand{\green}{\textcolor{black}}
\newcommand{\stella}{\texttt{stella}~}

\shorttitle{Turbulent impurity transport in W7-X}
\shortauthor{J. M. Garc\'ia-Rega\~na, M. Barnes, I. Calvo, F. I. Parra et al.}

\title{Turbulent impurity transport simulations in Wendelstein 7-X plasmas}

\author{J.~M.~Garc\'ia-Rega\~na\aff{1}\corresp{\email{jose.regana@ciemat.es}},
  M.~Barnes\aff{2},
  I.~Calvo\aff{1},
  F.~I.~Parra\aff{2},
  J.~Alcus\'on\aff{3},
  R.~Davies\aff{4},
  A.~Gonz\'alez-Jerez\aff{1},
  A.~Moll\'en\aff{5},
  E.~S\'anchez\aff{1},
  J.~L.~Velasco\aff{1},
  A.~Zocco\aff{3}}
\affiliation{\aff{1}Laboratorio Nacional de Fusi\'on, CIEMAT, Av. Complutense 28040, Spain
\aff{2}Rudolf Peierls Centre for Theoretical Physics, University of Oxford, Oxford OX1 3PU, UK
\aff{3}Max-Planck Institut f\"ur Plasmaphysik, Wendelsteinstrasse 1, 17491, Germany
\aff{4}York Plasma Institute, Department of Physics, University of York, Heslington, York YO10 5DD, UK 
\aff{5}Princeton Plasma Physics Laboratory, Princeton, NJ 08543-0451, USA}

\begin{document}

\maketitle

\begin{abstract}
A study of turbulent impurity transport by means of quasilinear
and nonlinear gyrokinetic simulations is presented for Wendelstein 7-X (W7-X).
The calculations have been carried out
with the recently developed gyrokinetic code \texttt{stella}. 
Different impurity species are considered in the presence
of various types of background instabilities: ITG, TEM and ETG modes 
for the quasilinear part of the work; ITG and TEM for the 
nonlinear results. While the quasilinear approach allows \blue{one} to draw qualitative conclusions about the sign or relative importance of the various contributions to the flux, the nonlinear simulations quantitatively determine the size of the turbulent flux and check the extent to which the quasilinear conclusions hold. Although the bulk of the nonlinear simulations are performed at trace impurity concentration, nonlinear simulations are also carried out at realistic effective charge values, in order to know to what degree the conclusions based on the  simulations performed \red{for trace impurities} can be extrapolated to realistic impurity concentrations. \blue{The presented results conclude that the turbulent radial impurity transport in W7-X is mainly dominated by ordinary diffusion, which is close to that measured during the recent W7-X experimental campaigns. It is also confirmed that thermo-diffusion adds a weak inward flux contribution and that, in the absence of impurity temperature and density gradients, ITG- and TEM-driven turbulence push the impurities inwards and outwards, respectively.}
\end{abstract}

\section{Introduction}

Impurity sources are inherent to the operation of present day fusion devices 
and will also be present in future reactors. 
Erosion from the first wall can release impurities to the plasma core, which 
can lead to the radiative collapse of the plasma if the impurity concentration becomes
sufficiently high.  
Impurities can also be intentionally introduced \blue{in the plasma} to access
the density and radiative conditions for divertor detachment, reducing the heat
loads over \green{the divertor} surface to tolerable levels.
In reactors, thermalized alpha particles will constitute 
the main impurity in the plasma core, and its removal
will be critical to avoid the dilution of
the D-T fuel. Impurities are also on the design basis of
different diagnostics of  bulk plasma properties, like
spectroscopy-based measurements of plasma flows, main ion temperature 
or radial electric fields.
For these reasons, substantial efforts have been devoted, in stellarator and tokamak 
experiments, theory and numerical simulations, to the identification of the
mechanisms that control impurity transport.

In stellarators, the concern for impurity accumulation arises from its observation in experiments \red{\citep{Burhenn_nf_49_2009, Hirsch_ppcf_50.5_053001_2008}, more severe in \textit{ion root} conditions (negative radial electric field) that standard neoclassical theory predicts when the main ion and electron temperatures are comparable. 
However, some scenarios have been identified too, that contradict that tendency, like 
the high-density H-mode (HDH) W7-AS plasmas \citep{McCormick_prl_89_015001_2002} 
and the \textit{impurity hole} LHD scenarios \citep{Ida_pop_16_056111_2009}.}
The existence of these scenarios has been the drive of a recent 
intense revision of neoclassical theory and numerical modeling, starting with the impact on impurity 
transport of the full neoclassical electric field, not
only radial but also tangential to the flux surfaces 
\citep{Garcia-Regana2013a, Regana_nf_57_056004_2017, Calvo_jpp_2018}; 
the role of the tangential components of the magnetic drift and the electric field has been rigorously formulated \citep{Calvo_ppcf_59_055014_2017} and numerically implemented in the recently released code 
\texttt{KNOSOS} \citep{Velasco_ppcf_2018, Velasco_jcp_2020}; 
\red{these advances have gone along with} more accurate treatments of 
collisions in self-consistent multispecies radially local simulations \citep{Mollen_ppcf_60_084001_2018};
the so-called \textit{mixed-collisionality-regime} (low collisional main ions and highly collisional impurity ions) 
has been uncovered with important implications regarding ion temperature screening 
\citep{Helander_PRL_118_155002, Calvo_NF_58_124005_2018, Buller_jpp_84_2018}; 
\blue{the importance of the classical transport for highly charged impurities has been reinvigorated in optimized stellarators \citep{Buller2019}};
finally, the first radially global neoclassical simulations including all these new neoclassical ingredients have been 
recently released \citep{Fujita_jpp_2020}. The outcome of these works has made evident that 
this broader neoclassical framework can introduce corrections of order unity in the impurity fluxes 
respect to the predictions of standard neoclassical theory, like those based on the drift kinetic equation solver DKES \citep{Hirshman_pf_29_2951_1986}.
\blue{However, such corrections can difficultly \green{explain} the order-of-magnitude discrepancy found in W7-X
between the experimentally measured diffusion coefficient of LBO-injected iron 
and that obtained with DKES \citep{Geiger_NF_59_046009_2019}}. Impurity confinement time scaling studies \citep{Langenberg2020} have also supported the hypothesis that the 
drive of impurity transport in W7-X plasmas has a significant turbulent component, resulting in the absence of impurity accumulation in most scenarios of the first operation phase, including 
those of high density with high likeliness of developing large ion-root electric fields \citep{Klinger_nf_59_112004_2019}.

With regard to impurity transport driven by gyrokinetic microturbulence, little 
work has been \blue{done} for stellarator geometry.
Among the few examples that have attempted to model it, the quasilinear 
analysis performed with the code GS2 in \citet{Mikkelsen_pop_21_082302_2014}
is one of the first examples available in the literature. 
Only very recently, nonlinear impurity transport simulations have also been 
carried out with the code GKV and reported in \citet{Nunami_pop_27_052501_2020}.
In both cases, the motivation was to capture the above-mentioned hollow 
impurity density profiles observed in LHD.
Apart from these numerical examples, 
some basic features of the quasilinear flux of impurities from gyrokinetic instabilities with $k_{\bot}\rho_i\lesssim 1$ \red{have} been analytically estimated in the collisionless 
electrostatic limit in \citet{Helander_ppcf_60_084006_2018} like, for example, the relative
size of the different diffusive and convective contributions to the flux or their signs. \blue{This work has been generalized including the effect of collisions \citep{Buller2020}, which are not considered in the present work}. \\
Therefore, the aim of the present work is building, by means
of linear and nonlinear gyrokinetic simulations, a first numerical characterization of 
the radial turbulent transport of impurities in W7-X plasmas. By doing this, \green{we pursue} to  alleviate the 
lack of numerical results for stellarators and to shed light on the interpretation of
W7-X experimental measurements.
The \red{analyses} that follow consider a set of selected impurities and  
bulk species gradients such that the \red{triggered} background instabilities are 
representative of ITG, TEM and ETG \red{modes}. All the numerical \red{work is} presented in 
section \ref{sec:numerical}, which is divided into three \red{subsections}. 
In the first of them, section \ref{sec:quasilinear}, the results presented are quasilinear and, through fast simulations that include ions, electrons and a single impurity at a trace concentration \red{level}, provide an overview 
of the relative weight, sign, mass, charge dependence, etc.~of
each diffusive or convective contribution to the turbulent particle
transport spectra for the selected impurities. 
Section \ref{sec:nonlinear} presents nonlinear simulations that, 
considering similar parameters than those employed for the quasilinear calculations, 
provide a quantitative evaluation of the actual size of diffusion and convection 
coefficients. Finally, the experimentally relevant situation of non-trace impurity content
is briefly discussed in section \ref{sec:nonlinear2}. 
All the calculations performed have been obtained with 
the newly developed stellarator gyrokinetic code \texttt{stella} \citep{Barnes_jcp_391_2019}. 
Finally, the conclusions are summarized 
in section \ref{sec:conclusions}.

\section{Numerical results}
\label{sec:numerical}

In the present section, the numerical results of turbulent impurity transport 
with the code \stella are presented 
and discussed. A complete description of the code can be found 
in \citep{Barnes_jcp_391_2019} but, for convenience, 
its main features are concisely summarized below.

\stella is a recently developed $\delta f$ code
whose current version solves, in the flux tube approximation, 
the gyrokinetic Vlasov and Poisson equations for an arbitrary number 
of species. The magnetic geometry can be specified either by the set of Miller's 
parameters for a local tokamak equilibrium or by a 3D equilibrium 
generated with \texttt{VMEC}, which has been the option considered for  
the simulations carried out in this paper. 
The spatial coordinates that the code uses for stellarator simulations are: 
the flux surface label $x=a\sqrt{s}$ (commonly denoted by $r$), with $a$ the minor radius of the device and 
$s=\psi_t/\psi_{t,\mathrm{LCFS}}$ the toroidal magnetic flux normalized 
to its value at the last closed flux surface; 
the magnetic field line label $y=a\sqrt{s_0}\alpha$, a rescaled version of the Clebsch angle $\alpha=\theta^*-\iota\zeta$, 
with $\theta^*$ and $\zeta$ the poloidal and toroidal, respectively, PEST flux coordinates \green{\citep{Grimm_jcp_49_1983}}, 
$\iota$ the rotational transform and $s_0$ the 
value of the flux surface label around which the flux tube is centered; 
the parallel coordinate $z=\zeta$. The velocity coordinates
are the magnetic moment $\mu$ and the parallel velocity $v_{\|}$.
The crucial feature of the algorithm employed by \stella
to solve the gyrokinetic equation is the mixed implicit-explicit 
treatment of its different terms. In particular, 
a splitting of the Vlasov operator is applied, and the pieces containing
the parallel streaming and acceleration are treated implicitly. 
For electrons, these pieces
scale up to a factor of order $\sqrt{m_i/m_e}$ (with $m_i$ and $m_e$ 
the main ion and electron mass, respectively)
respect to all other terms in the gyrokinetic equation, imposing  
in fully explicit time-advance schemes a severe restriction to the time step
size, tighter at lower perpendicular wavenumber $k_{\bot}$. 
The mixed implicit-explicit algorithm employed by \stella relaxes
this constraint on the time-step and allows \green{on to include kinetic electrons in multispecies simulations} with practically no increase of computational cost
apart from the required to loop over more species.

Returning to the impurity transport problem, the possibility of including kinetic electrons with practically no need of decreasing the time step, has made it possible to address with multiple nonlinear simulations 
the quantitative characterization of the transport of impurities 
under both ion- and electron-driven background turbulence in W7-X geometry.
This is the \textit{raison d'\^etre} of the present 
work, and the obtained results, discussed in detail in 
section \ref{sec:nonlinear}, its main achievement. However, 
given the very few stellarator references addressing this problem 
even on a quasilinear fashion, section \ref{sec:quasilinear} is dedicated to a quasilinear
analysis that precedes the nonlinear treatment of section \ref{sec:nonlinear}. 
To what extent the conclusions drawn from the presented quasilinear calculations follow analytical quasilinear theory predictions \citep{Helander_ppcf_60_084006_2018} and hold in light of nonlinear results will be briefly commented.

Sections \ref{sec:quasilinear} and \ref{sec:nonlinear} consider impurities 
at trace concentration, which allows \green{us} to assume that transport coefficients
are independent of the impurity density and temperature gradients and 
to express the \red{turbulent radial impurity flux} as:

\begin{equation}
\Gamma_Z=-n_Z\left(D_{Z1}\frac{\text{d}\ln n_Z}{\text{d} r}+D_{Z2}\frac{\text{d}\ln T_Z}{\text{d} r}+C_Z\right)
\label{eq:transport_law}
\end{equation}
with $D_{Z1}$ the impurity diffusion coefficient, $D_{Z2}$ the thermo-diffusion 
coefficient, and $C_Z$ the flux in the absence of impurity density and temperature gradients, which includes the contribution from the curvature pinch and the flux arising from the \green{accelleration of impurities due to the turbulent parallel electric field}\footnote{Note that, regardless the terminology, only the coefficient multiplying the impurity density gradient, $D_{Z1}$, is a \textit{diffusive} term and the rest 
are \textit{convective} terms. In other words, following the widely employed expression
$\Gamma_Z/n_z=-D\text{d}\ln n_Z/\text{d}r + V$, see e.g.~\cite{Burhenn_nf_49_2009},
\red{$V=-(D_{Z2}\dd\ln T_Z/\dd r +C_Z)$} corresponds to the commonly named as convection 
velocity and $D=D_{Z1}$ to the so-called diffusion coefficient.}.  
In expression (\ref{eq:transport_law})
$n_Z$ and $T_Z$ are the impurity density and temperature, respectively. 
\red{Finally, in section \ref{sec:nonlinear2} the question about the dependence that the transport coefficients develop at non-trace impurity concentration is addressed. In particular, the impurity flux scaling with the impurity density gradient at $Z_{\text{eff}}=2$ is \green{investigated} by means of nonlinear simulations, in order to \green{determine} whether the conclusions drawn assuming the trace limit can be extrapolated to more realistic plasma conditions.}

All simulations, linear and nonlinear, at trace and non-trace impurity 
content, have in common: the magnetic geometry, which is the 
\textit{standard} W7-X configuration (see \citet{Geiger_ppcf_57_1_2015}
for an overview of the W7-X configuration space);
the flux tube chosen, $\alpha=0$ \blue{, as it is usually found to be the most unstable flux tube in W7-X (see \citet{Helander2012} for a discussion about the localization of the turbulent fluctuations of the electrostatic potential along this flux tube)}; the main ion species, hydrogen, 
and the chosen flux surface, $\sqrt{s_0}=0.49$. 
Other parameters, specific of the type of simulations performed, are given in the corresponding 
section.

\subsection{Linear stability and quasilinear impurity transport analysis}
\label{sec:quasilinear}

\begin{table}
\begin{center}
\begin{tabular}{ccccc}
& $a/L_{T_i}$ & $a/L_{T_e}$ & $a/L_{n_i}=a/L_{n_e}$ & $T_e/T_i$ \\
\hline
ITG & $4.0$ & $0.0$ & $0.0$ & 1.0 \\
TEM & $0.0$ & $0.0$ & $4.0$ & 1.0 \\
ETG & $0.0$ & $4.0$ & $0.0$ & 1.0 \\
\hline
Species & \multicolumn{3}{c}{Ar$^{16+}$, Mo$^{16+}$, W$^{16+}$, W$^{30+}$, W$^{44+}$}\\
\hline
\end{tabular}
\end{center}
\caption{Normalized gradients, electron to ion temperature ratio, and selected
impurities considered for the quasilinear transport study.}\label{table:quasilinear}
\end{table}

How the impurity transport is affected by the driven gyrokinetic electrostatic instabilities of a set of specific LHD \textit{impurity hole} discharges can be found in \citet{Mikkelsen_pop_21_082302_2014}. However, a similar analysis is not reported, to our knowledge, for W7-X geometry, which motivates \green{us} to perform a quasilinear characterization of the \red{turbulent} impurity transport in this device prior to turning to the fully nonlinear treatment in section \ref{sec:nonlinear}.
In addition, recent work by \citet{Helander_ppcf_60_084006_2018}
has analytically deduced some qualitative features of the quasilinear transport coefficient 
of impurities, which can be contrasted with the presented quasilinear numerical estimations.

The selected parameters and impurity species for the quasilinear study are summarized in 
table \ref{table:quasilinear}.
The gradients of the bulk species have been set such that hybrid instabilies were discarded. That is, 
$\left\{a/L_{T_i}, a/L_{T_e}, a/L_{n_e}\right\}=\left\{4, 0, 0\right\}$
has been set for the ITG driven instability, $\left\{a/L_{T_i}, a/L_{T_e}, a/L_{n_e}\right\}=\left\{0, 0, 4\right\}$
for density gradient driven TEMs, and 
$\left\{a/L_{T_i}, a/L_{T_e}, a/L_{n_e}\right\}=\left\{0, 4, 0\right\}$
for ETG modes\footnote{\green{Note that the label of the instability driven solely by the electron temperature gradient as ETG has been taken for practical purposes, in order to ease the discussion about the impurity particle transport produced by types of turbulence driven each by the gradient of one single plasma parameter. This labeling obviates the fact that trappped electron modes can also be driven by the electron temperature gradient, see \citep{Proll2013} for a discussion about the characteristics of TEMs in stellarators.}}. 
For all cases the flux tube has been extended three turns poloidally,  
the wavenumber along the radial direction has been set 
to $k_x=0$, and the wavenumber along the binormal direction, $k_y$, has been scanned. 
All simulations have been performed with kinetic main ions, electrons
and a single impurity species at a trace concentration.
The set of selected impurities have included Ar$^{16+}$, Mo$^{16+}$, W$^{16+}$, W$^{30+}$ and W$^{44+}$.

\begin{figure}
  \begin{center}
  \includegraphics[height=5.5cm]{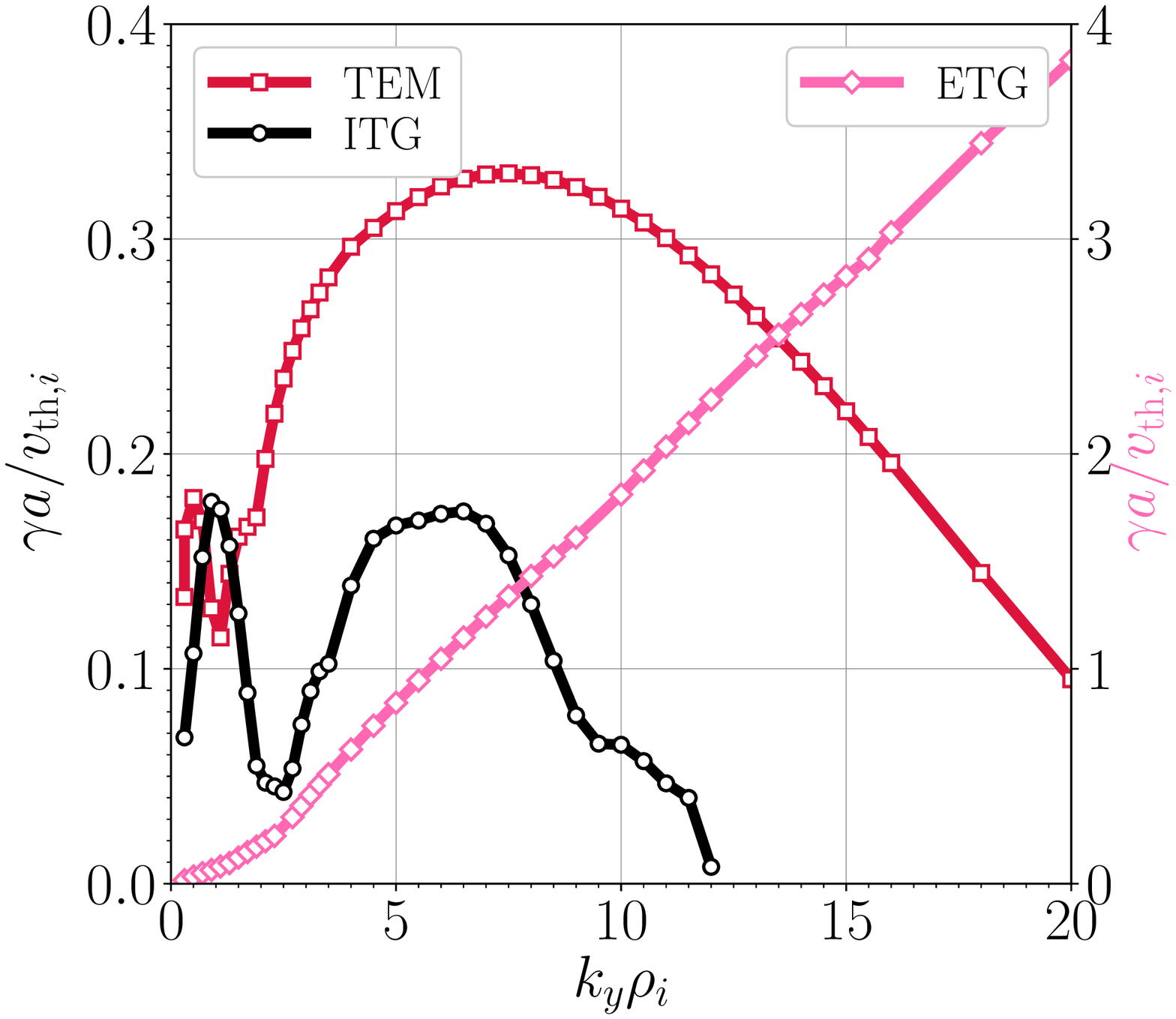}
  \includegraphics[height=5.5cm]{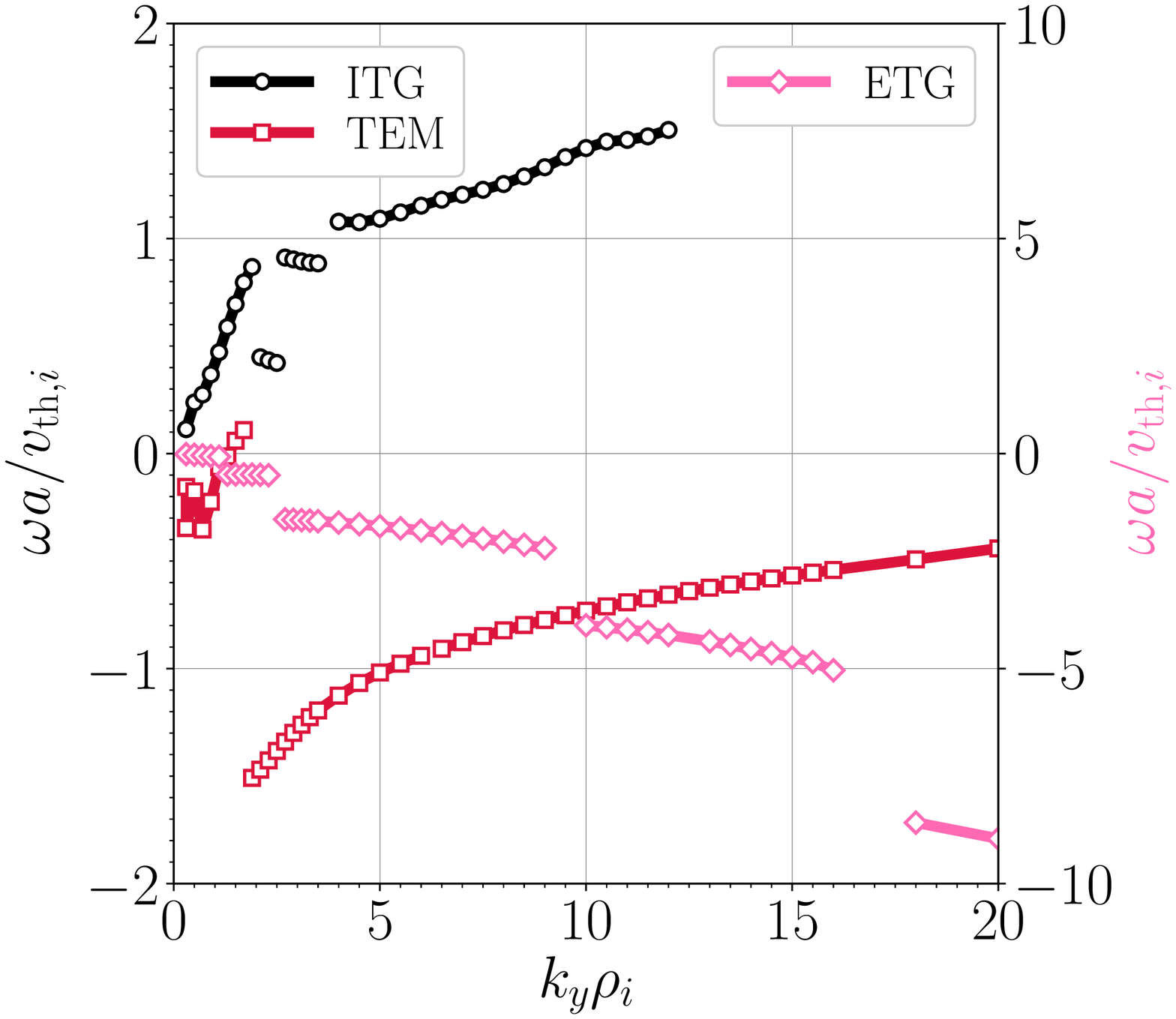}
  \end{center}
  \caption{Normalized growth rate (left) and frequency (right) as a function of the 
  normalized   binormal wavenumber $k_y$ for the three sets of background unstable conditions considered 
  for the quasilinear impurity transport study, 
  namely, ITG driven by $a/L_{T_{i}}=4.0$ (circles), TEM driven by $a/L_{n_{e}}=a/L_{n_{i}}=4.0$ 
  (squares) and ETG driven $a/L_{T_{e}}=4.0$ with (diamonds). Here, the normalization 
  uses the ion Larmor radius, $\rho_i$, the ion thermal speed, $v_{\text{th},i}$, 
  and the effective minor radius, $a$.}
  \label{fig:linspectra}
\end{figure}


The spectra of the growth rate, $\gamma$, and frequency, $\omega$, for the three different 
linear instabilities simulated are represented in fig.~\ref{fig:linspectra} (left) and
fig.~\ref{fig:linspectra} (right), respectively.  
It is observed that the ITG-driven instability features a double peak structure and 
extends over a considerably broad $k_y$ range up to $k_{y}\rho_i\approx 12$. 
However, the fastest growing mode is located 
at $k_{y}\rho_i\approx 1$. Changes in the dominant eigenmode 
can be inferred from the discontinuous spectrum of the frequency. The sign
of the frequency indicates that the modes rotate in the ion diamagnetic direction 
for all $k_y$. On the other hand, the density gradient driven TEM is found to be more
unstable that the ITG, with the fastest growing mode of the former 
featuring a factor two larger growth rate than that of the latter. The fastest growing mode 
is located at $k_y\rho_i\approx 7.5$, although the 
instability extends beyond $k_y\rho_i=20$. The sign of the frequency indicates that the mode can rotate both in the ion-diamagnetic direction
for the low $k_y$ part of the spectrum, and in electron diamagnetic direction at moderate and high $k_y$. Finally, the ETG-driven instability 
shows a monotonically increase of the growth rate towards electron Larmor scales, not covered on the simulated range of $k_y$, where the most unstable $k_y$ is expected to be located. Note, though, the large value of the growth rate (referred to the right y-axis) that the ETG-driven instability develops at scales of a few ion Larmor radius. The frequency, in this case, shows that the mode rotates in the electron diamagnetic direction and that different branches, presumably dominated by a different eigenmode, are encountered, as the discontinuous frequency pattern points out.

\begin{figure}
	\includegraphics[width=0.32\textwidth]{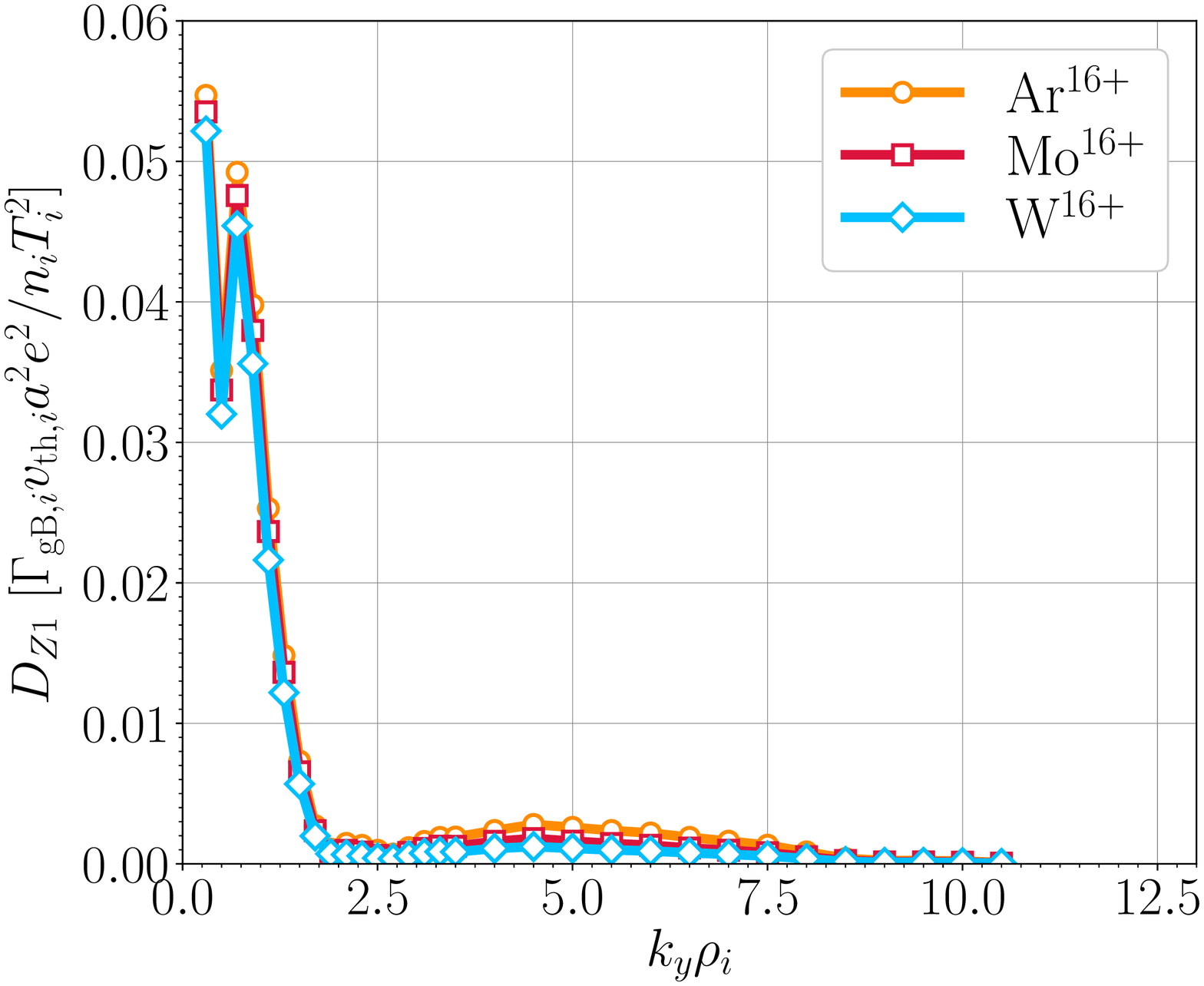}
	\includegraphics[width=0.32\textwidth]{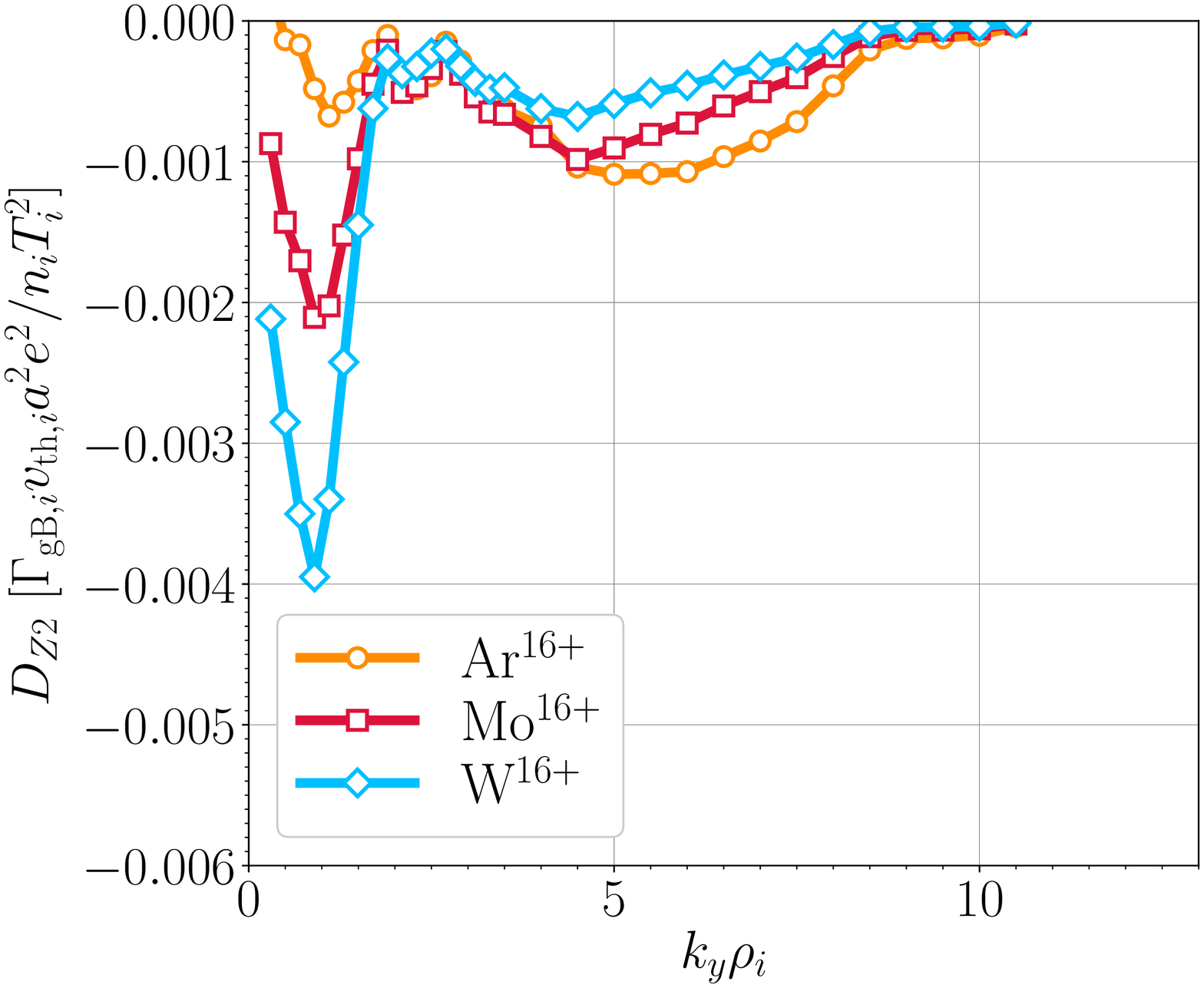}
	\includegraphics[width=0.32\textwidth]{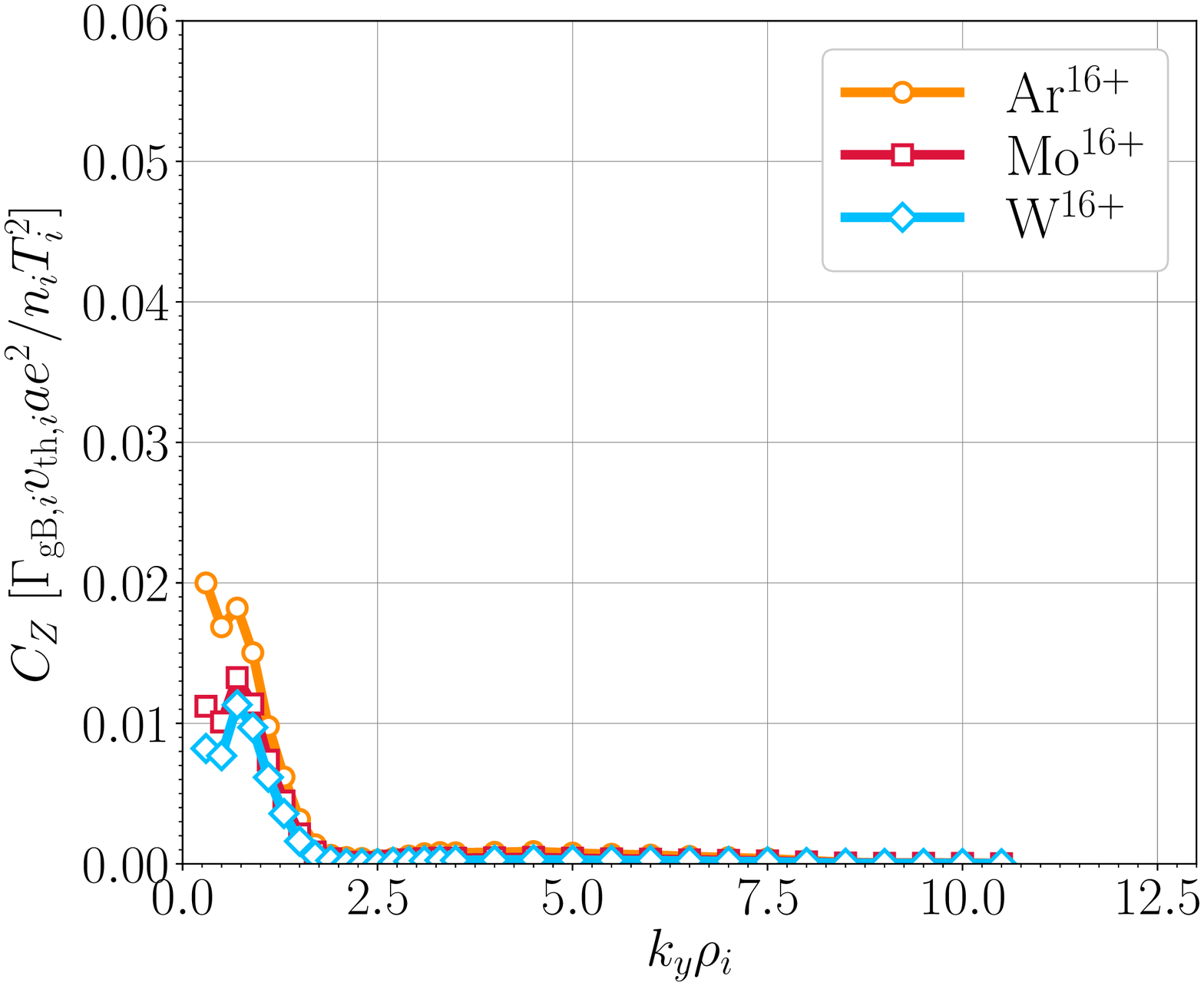}\\
	\includegraphics[width=0.32\textwidth]{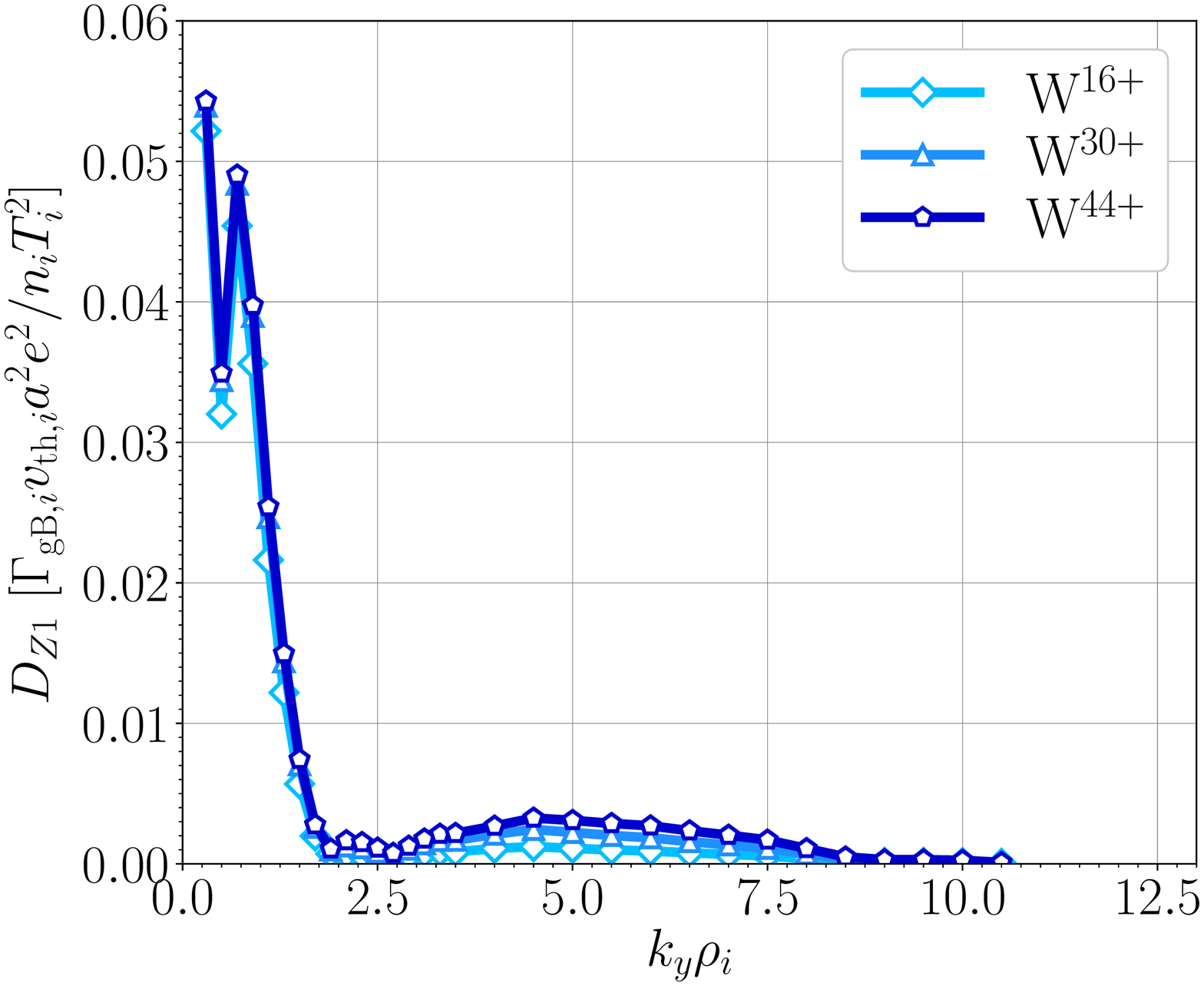}
	\includegraphics[width=0.32\textwidth]{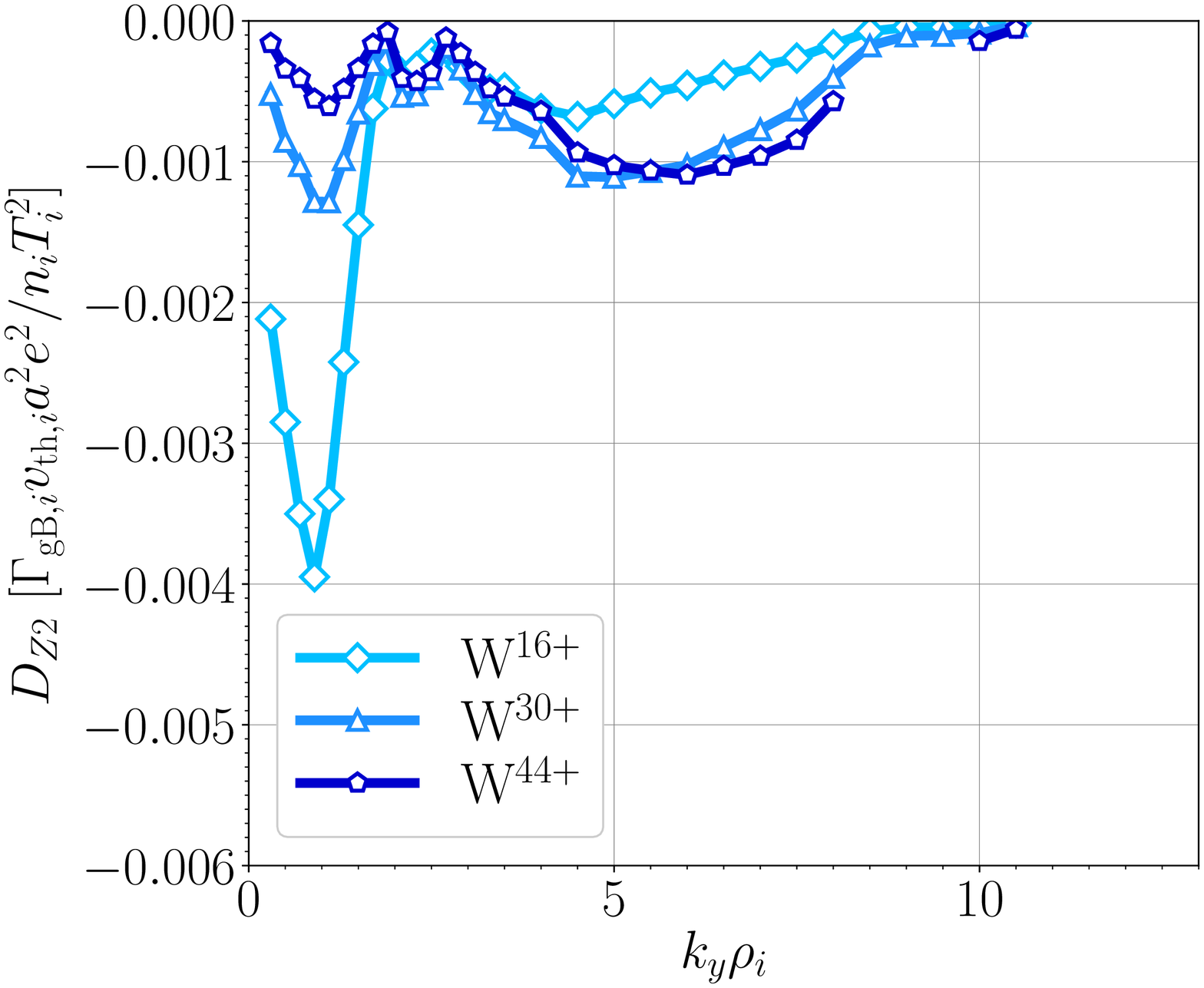}
	\includegraphics[width=0.32\textwidth]{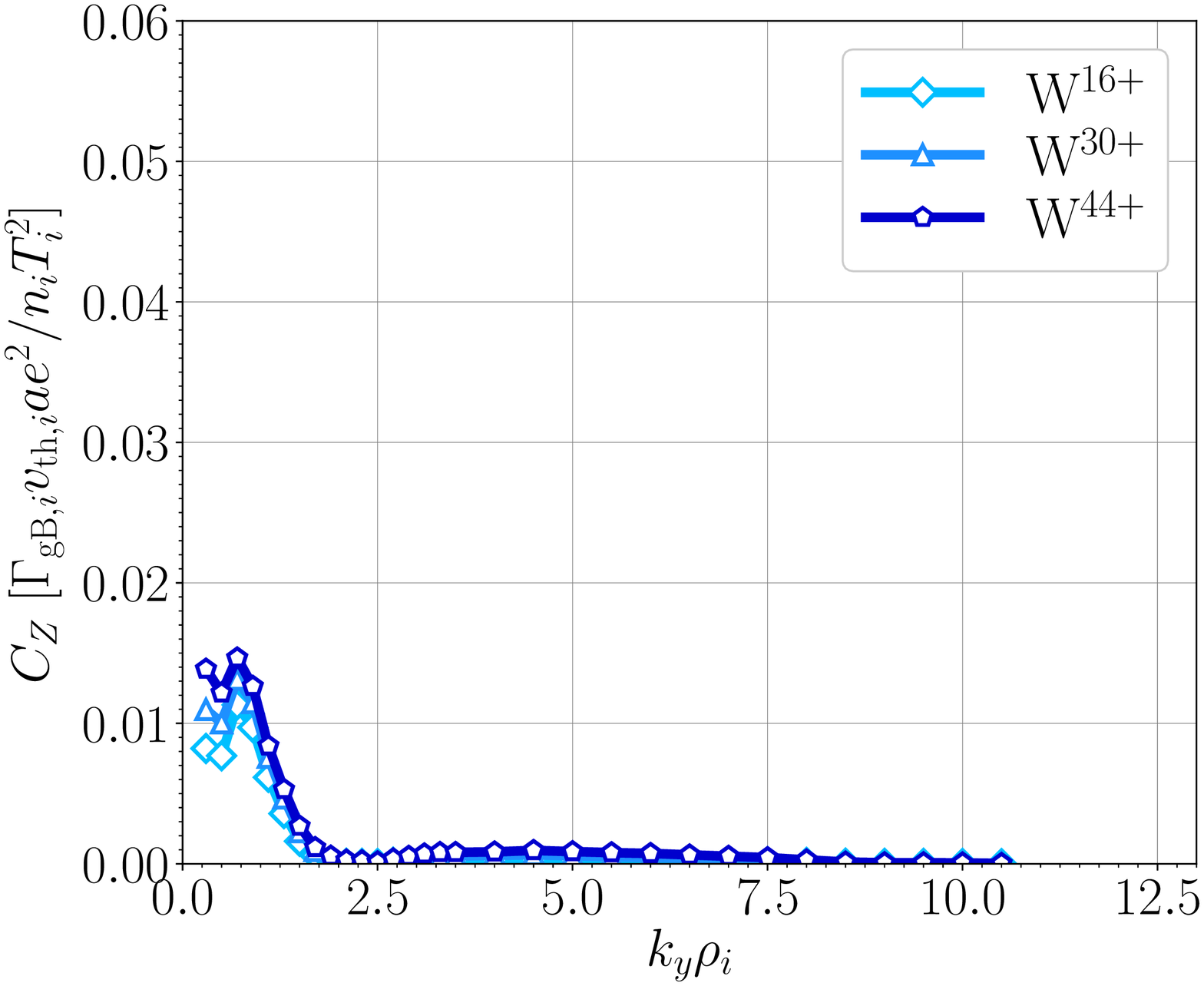}
	\caption{For the ITG case (see table \ref{table:quasilinear}), $k_y$-spectra of the diffusion coefficient $D_{Z1}$ (left), thermo-diffusion coefficient $D_{Z2}$ (center) and pinch in the absence of gradients $C_{Z}$ (right) for the different impurities, for a mass (top row) and charge (bottom row) scan. \blue{The values are given in normalized units, with $\Gamma_{\text{gB},i}$ the gyro-Bohm ion particle flux, $v_{\text{th},i}$ the ion thermal speed, $e$ the unit charge, $a$ is the effective minor radius and $n_i$ and $T_i$ the main ion density and temperature, respectively.}}
	\label{fig:qlitg}
\end{figure}

Returning to the question about the impurity transport driven by the above-mentioned instabilities, we have followed the same approach as in \cite{Mikkelsen_pop_21_082302_2014}. \red{Given a mode with wavenumbers $k_x$ and $k_y$, the linear Vlasov-Poisson gyrokinetic system of equations is solved for each simulated time step, and the gyroaveraged impurity distribution $g_Z(k_x, k_y,z,v_{\|}, \mu, t)$ and electrostatic potential $\varphi(k_x, k_y,z, t)$ are obtained. From these two quantities, the flux surface averaged impurity flux, $\Gamma_Z(k_x,k_y, t)$, is computed. Note that, once the instability has been triggered the electrostatic potential $\varphi$ and, consequently, $\Gamma_Z$ \green{grows} exponentially. However, a quasilinear mixing-length estimate of the flux,}
\begin{equation}
\red{\Gamma^{ql}_{Z}(k_x, k_y, t)=\frac{\Gamma_{Z}(k_x, k_y, t)\gamma(k_x, k_y, t)}
{n_Z\left<\varphi^2(k_x, k_y,z, t)\right>k_\bot^2(k_x,k_y)},}
\label{eq:ql}
\end{equation} 
\red{can be defined, so that a well converged quantity is obtained once the growth rate  is stabilized. In this expression $\left<...\right>$ denotes the flux surface average operator and $k_\bot=k_x\nabla x+k_y\nabla y$. 
Considering $k_x=0$ for all simulations, for each impurity species in the presence of a background instability, the $k_y$-spectrum of the quasilinear flux has been extracted at the last simulated time step. This process has been repeated with three different pairs of impurity density and temperature gradients, in order to obtain from each impurity species embedded in a different type of instabilies the spectra of the three transport coefficients}. 

For the ITG instability, the spectra of the diffusion coefficient, $D_{Z1}$, thermo-diffusion coefficient, $D_{Z2}$, and \blue{the impurity flux in the absence of impurity density and temperature gradients}, $C_Z$, are displayed on the left, center and right columns of fig.~\ref{fig:qlitg}, respectively. While the top row shows the results for the selected impurities with different mass, the bottom row does the same for the impurities with different charges. In first place, $D_{Z1}$ results to be roughly one order of magnitude larger than $D_{Z2}$, each having a different sign. That is, while diffusion drives impurities downhill the density gradient, thermo-diffusion would add an inward convection contribution, assuming peaked impurity temperature profile. In any case, this contribution seems very weak. Another inward contribution to the flux arises \blue{at vanishing impurity density and temperature gradients}, which, however, also seems comparatively small compared to the size of $D_{Z1}$. The spectra of the three transport coefficients show that most contributions to the total flux comes from the lowest part of the spectrum, from $k_y\rho_i\lesssim 1.5$. Finally, no significant dependence on the impurity charge or mass is observed for $D_{Z1}$ and $C_Z$. On the other had, \blue{the size of the weak $D_{Z2}$ is larger with} increasing mass and decreasing charge.

\begin{figure}
	\includegraphics[width=0.32\textwidth]{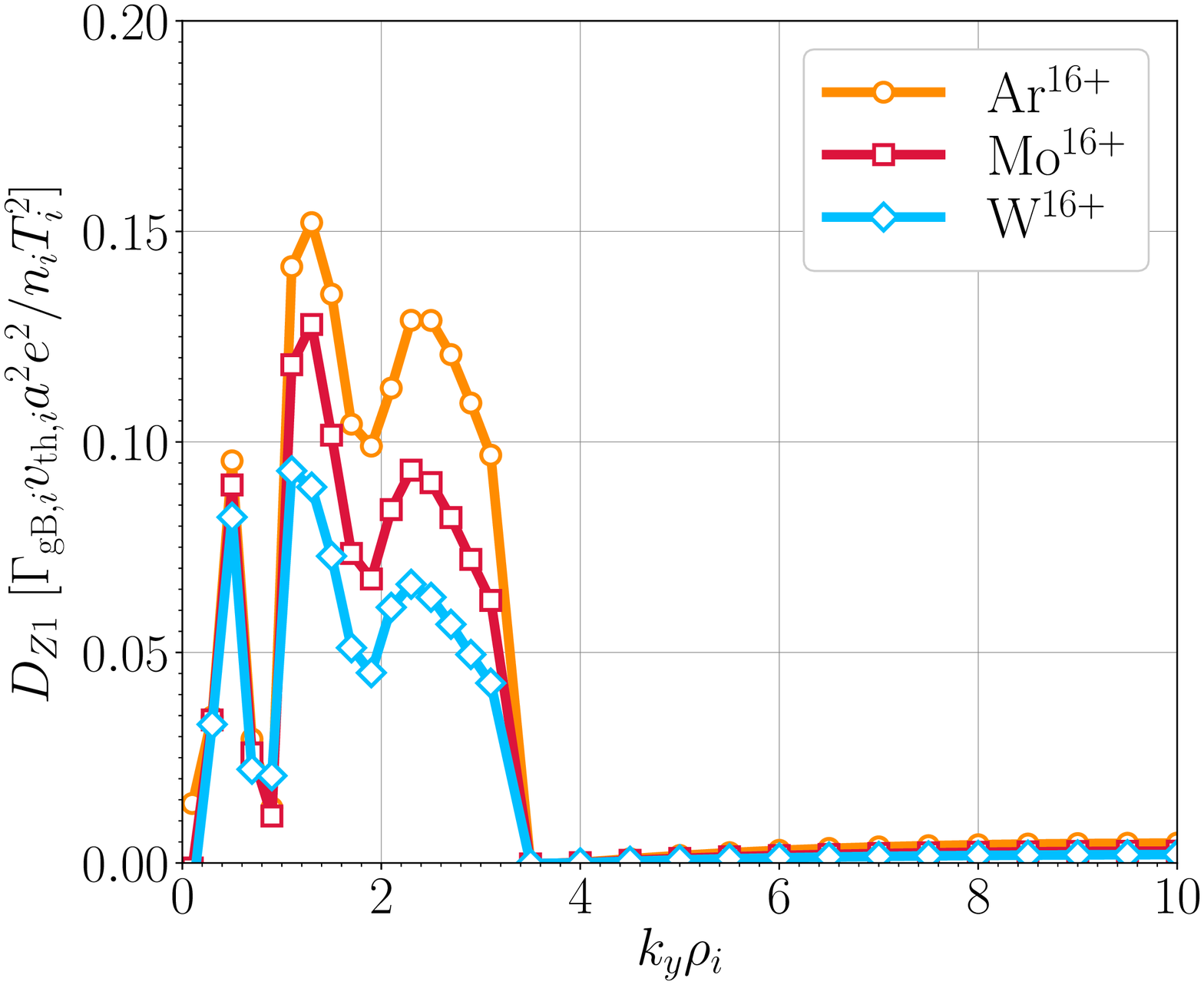}
	\includegraphics[width=0.32\textwidth]{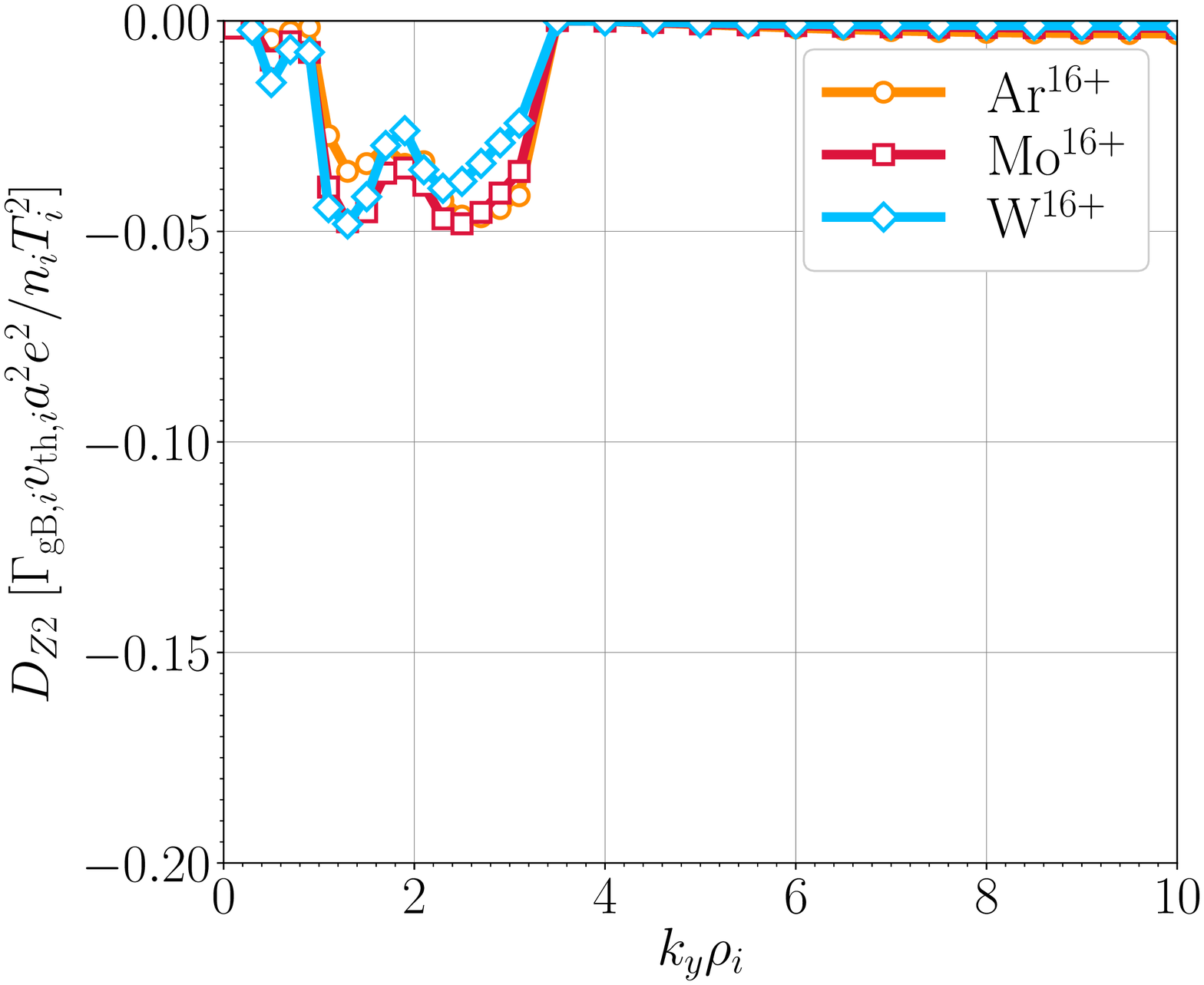}
	\includegraphics[width=0.32\textwidth]{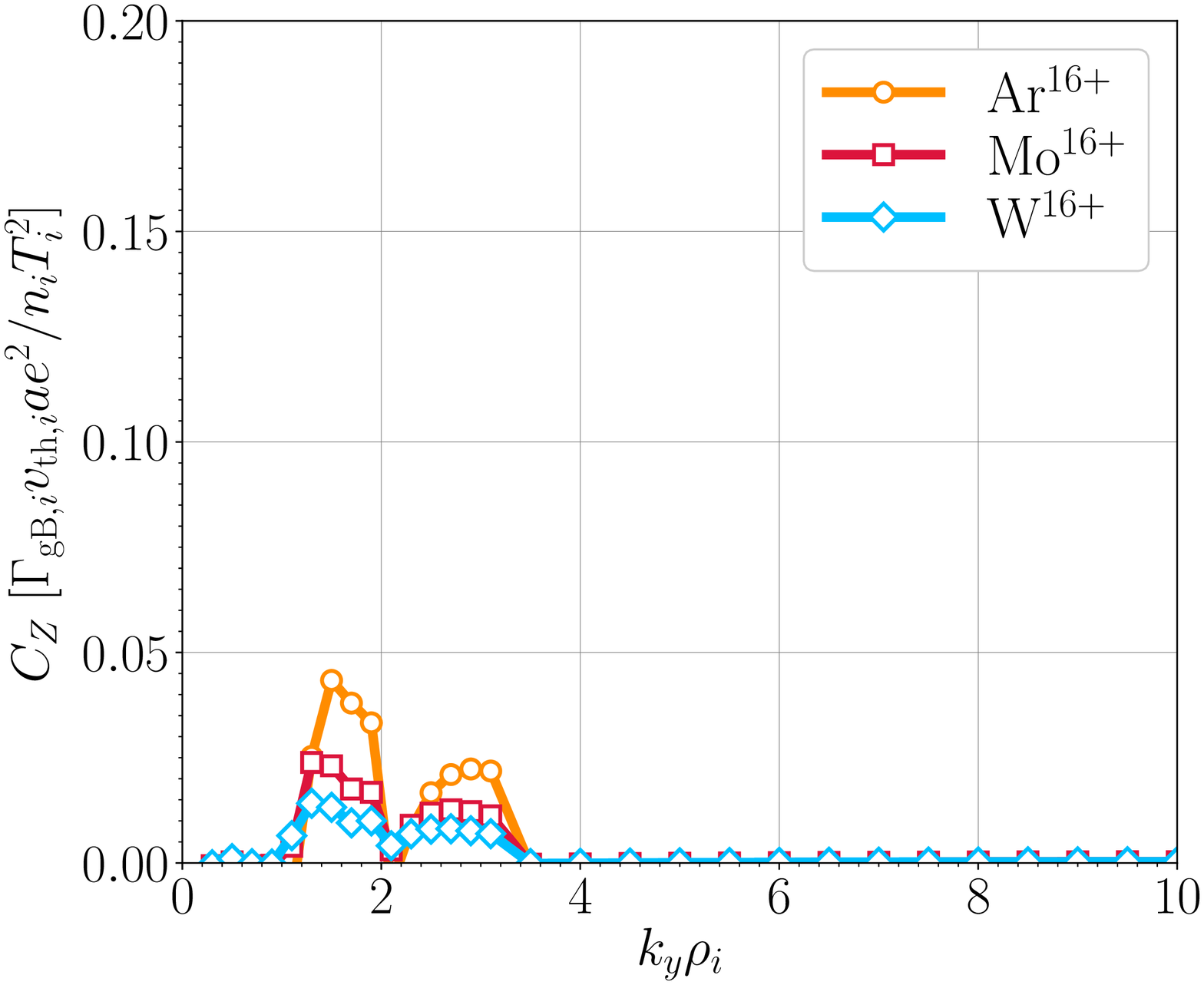}\\
	\includegraphics[width=0.32\textwidth]{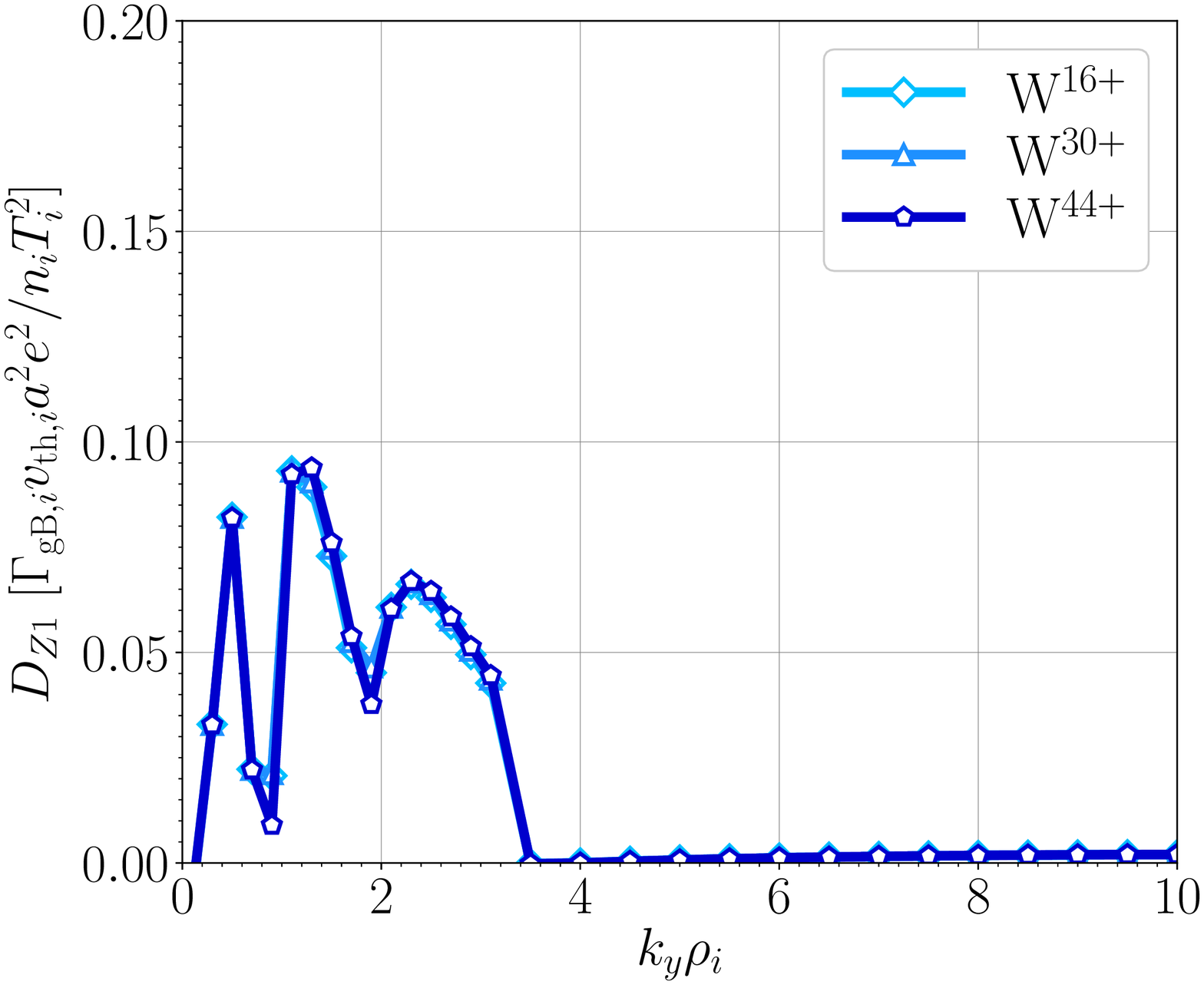}
	\includegraphics[width=0.32\textwidth]{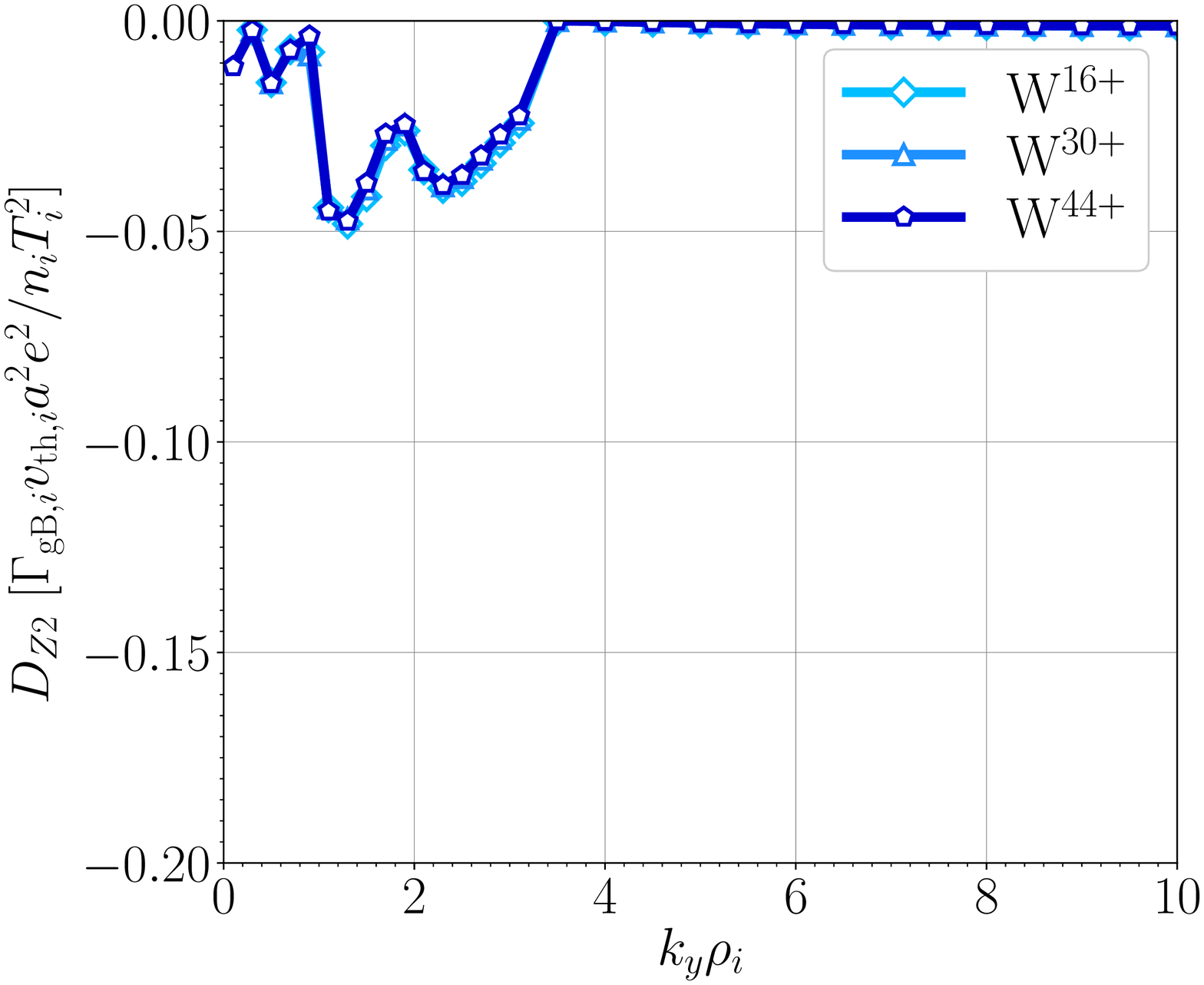}
	\includegraphics[width=0.32\textwidth]{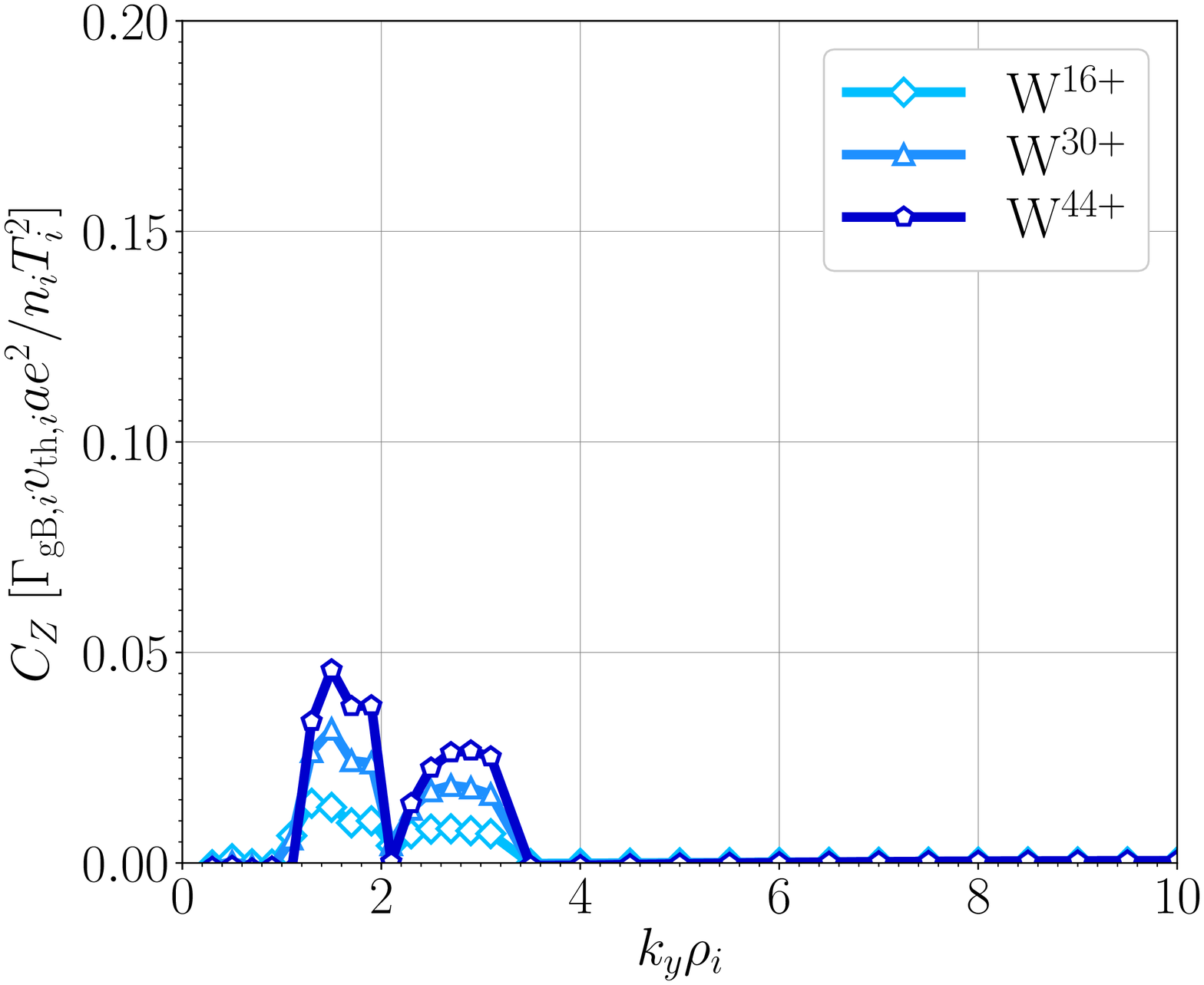}
	\caption{For the TEM case, see table \ref{table:quasilinear}, ky-spectra of the diffusion coefficient $D_{Z1}$ (left), thermo-diffusion coefficient $D_{Z2}$ (center) and pinch in the absence of gradients $C_{Z}$ (right) for the different impurities, for a mass (top row) and charge (bottom row) scan. }
	\label{fig:qltem}
\end{figure}

For the impurity transport coefficient driven by TEM instability, the corresponding results are shown in fig.\ref{fig:qltem}. In general, the transport coefficients follow the same trends as those observed in the ITG case. The diffusion coefficient is in absolute value larger than the thermo-diffusion, and the sign of each of them is the same as for the ITG instability. However, the difference between $D_{Z1}$ and $D_{Z2}$ is a factor of three while in the ITG case they differed \blue{by} roughly one order of magnitude. Furthermore, the strength of $D_{Z1}$ in this case is enhanced with respect to the ITG mode, possibly due to the \blue{more} unstable character of this TEM, see fig.~\ref{fig:linspectra}. Furthermore, the three \blue{coefficients} exhibit broader $k_y$-spectra than in the ITG case. Regarding the dependence of the coefficients on the mass or the charge, it is observed that $D_{Z1}$ somewhat depends on the charge and that $C_Z$ also depends on the charge and the mass.

\begin{figure}
	\includegraphics[width=0.32\textwidth]{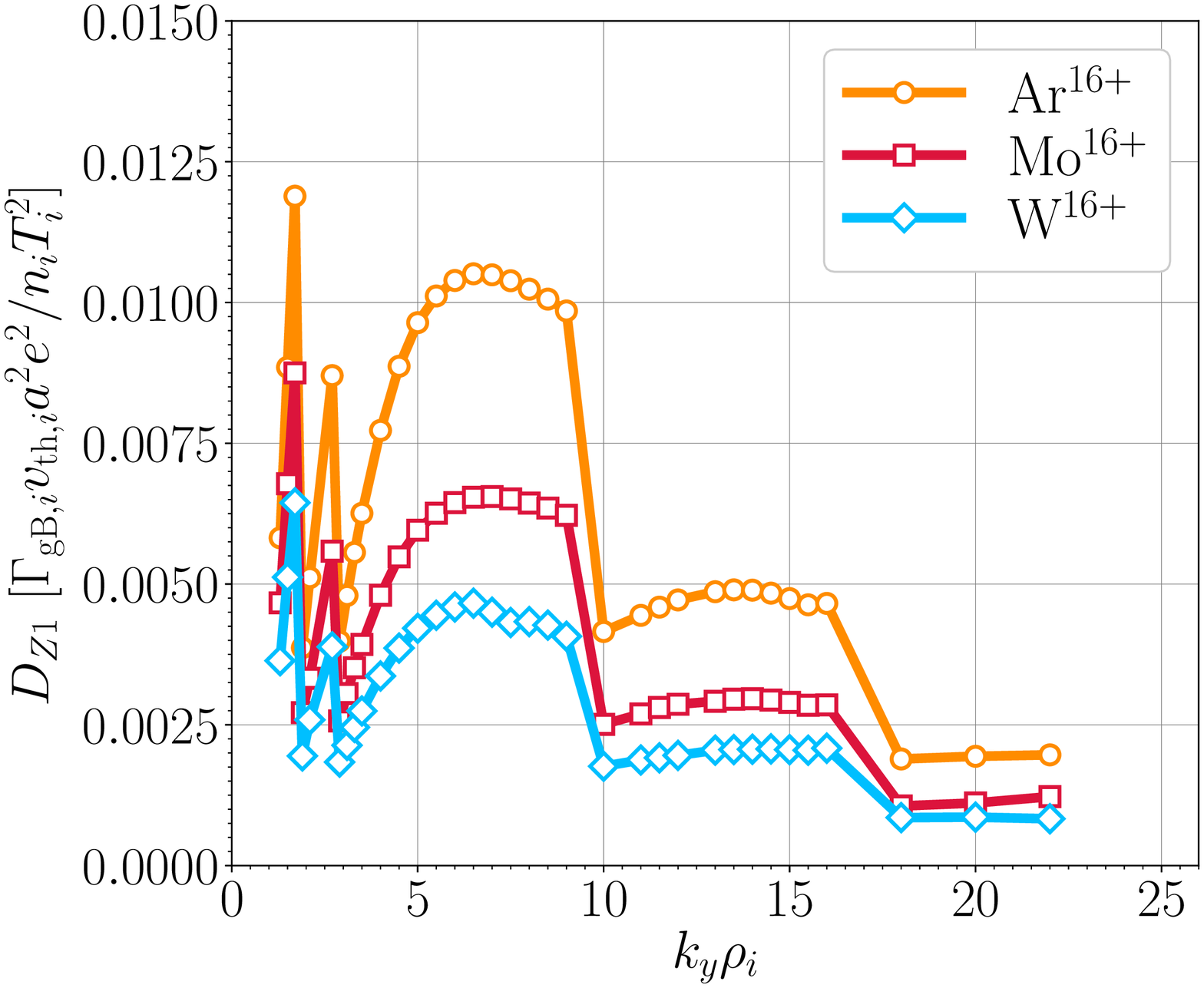}
	\includegraphics[width=0.32\textwidth]{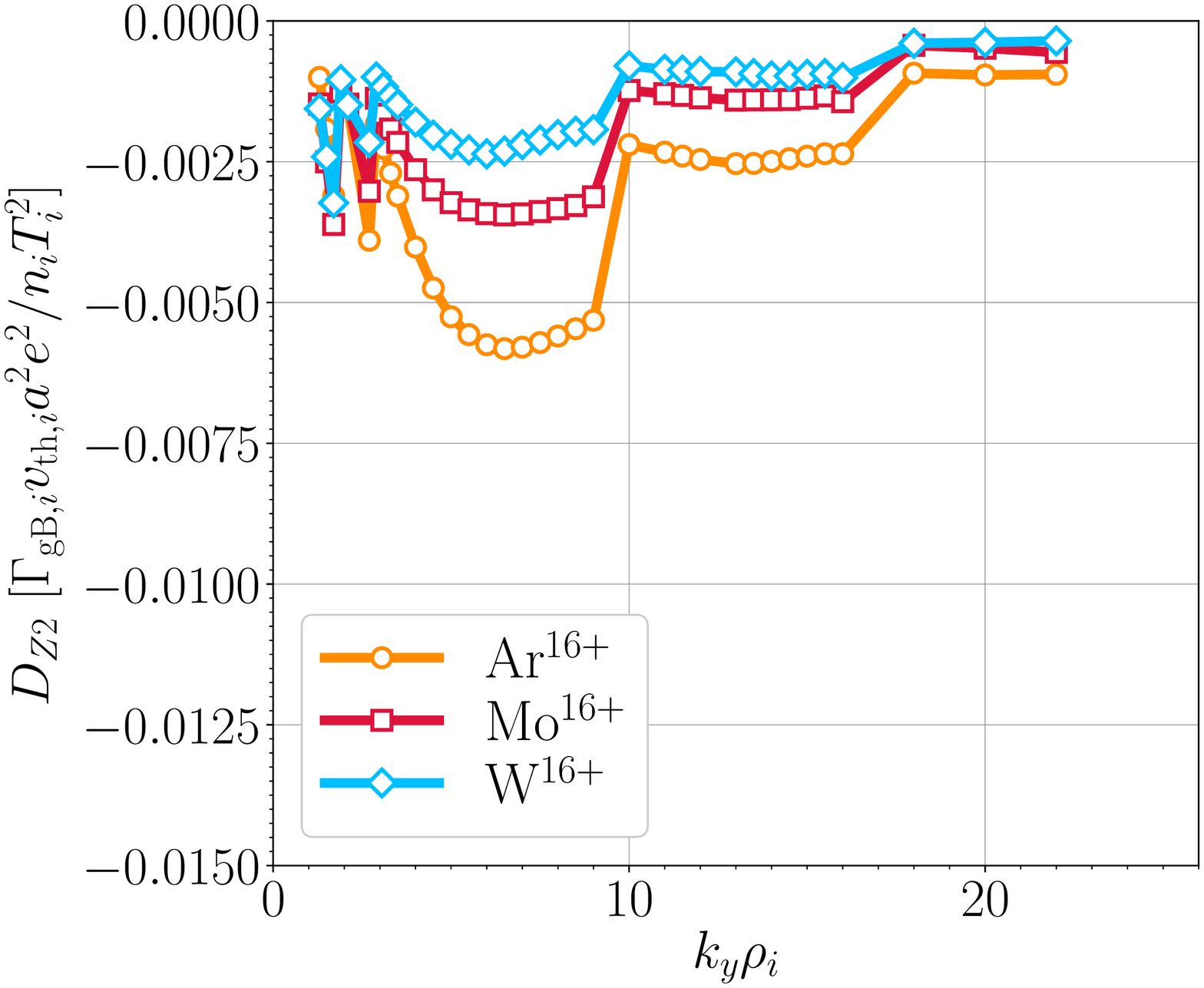}
	\includegraphics[width=0.32\textwidth]{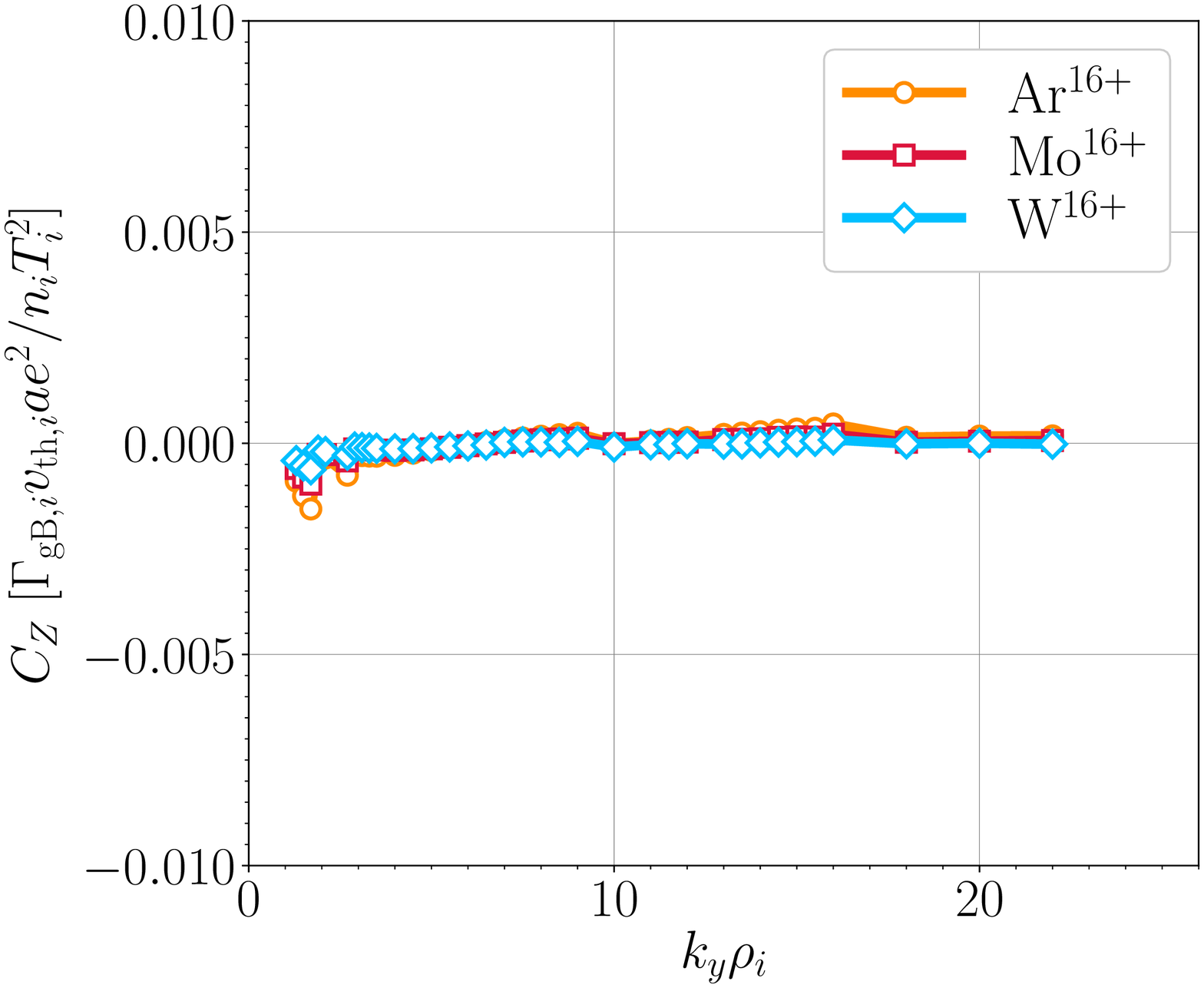}\\
	\includegraphics[width=0.32\textwidth]{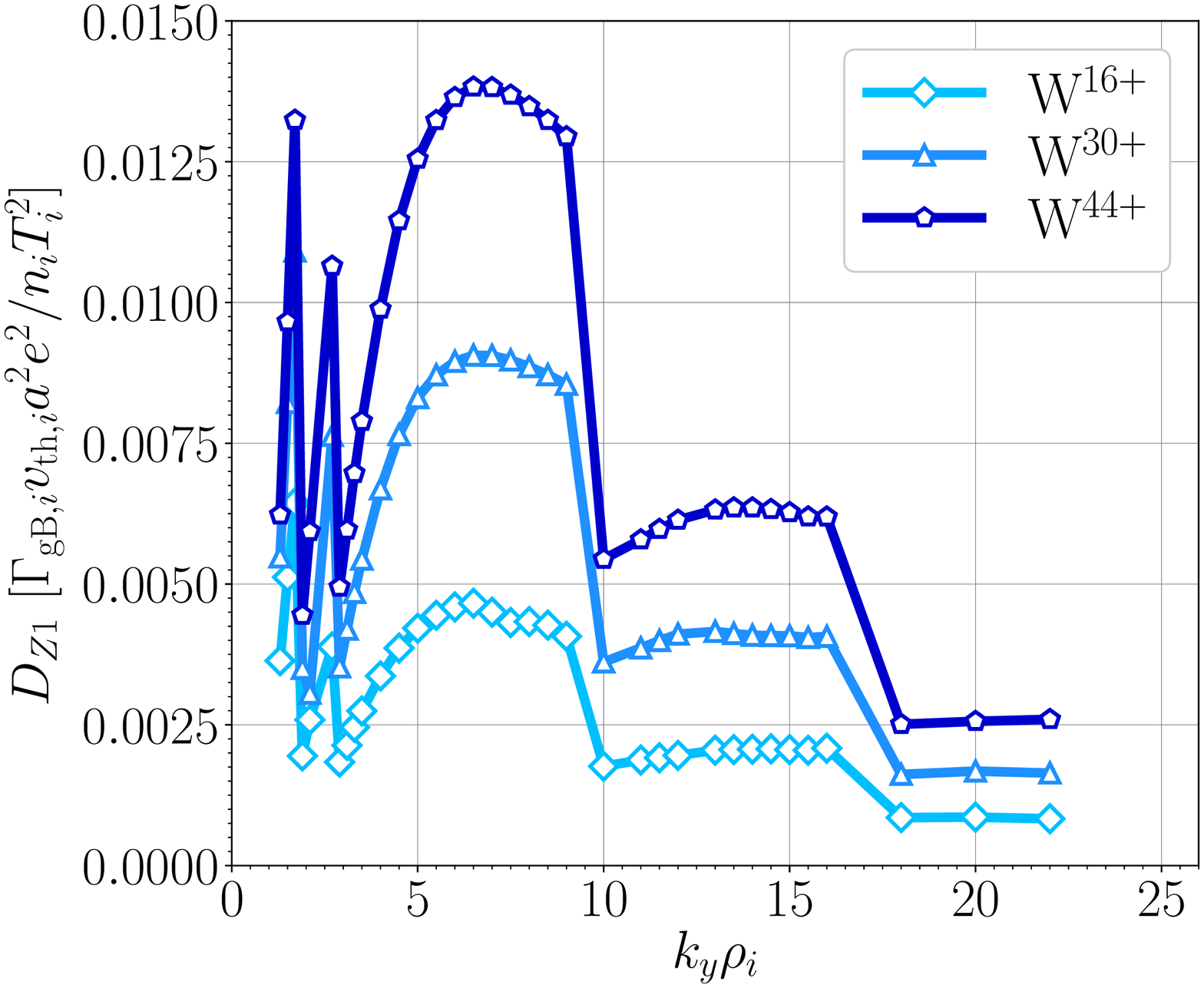}
	\includegraphics[width=0.32\textwidth]{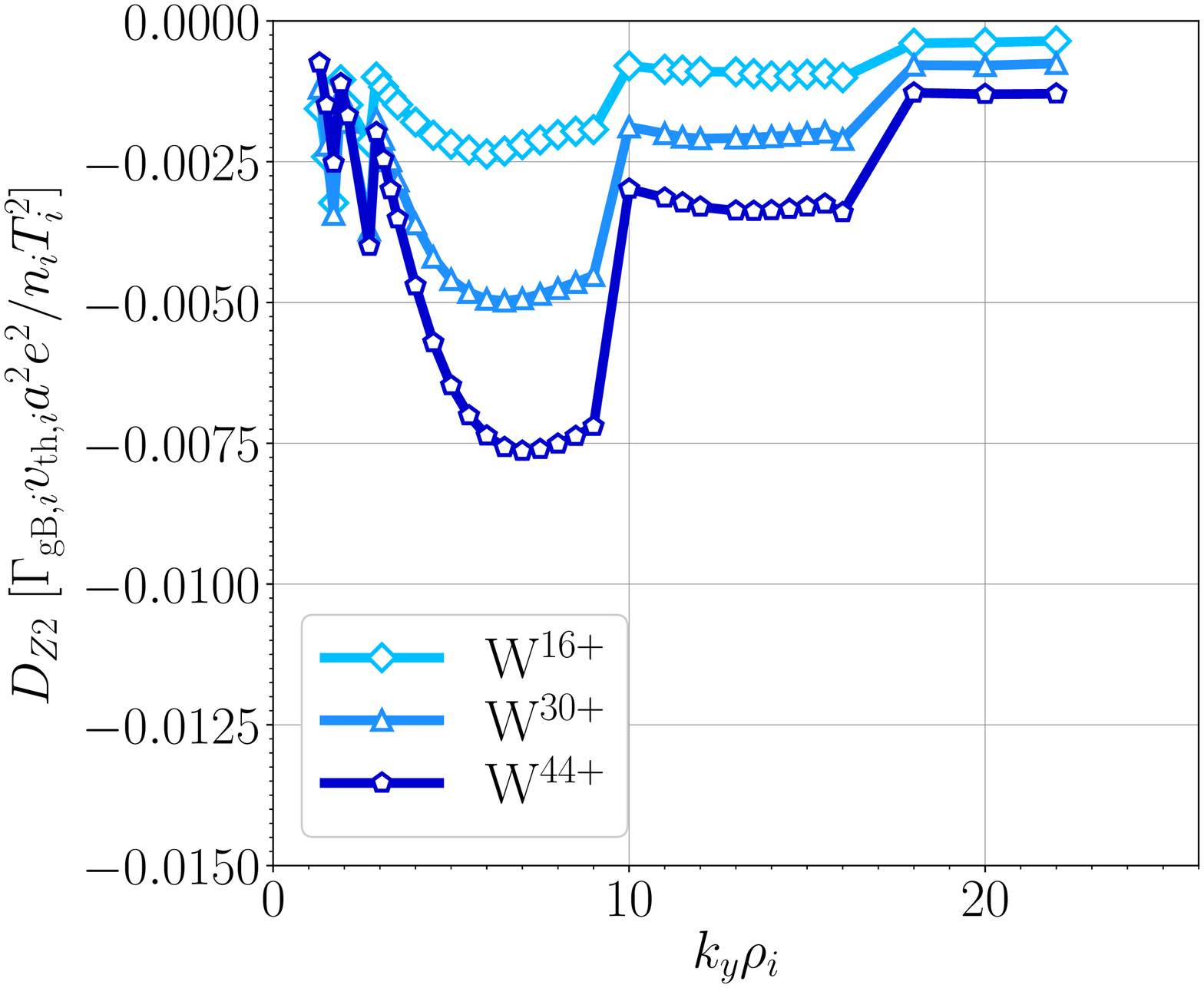}
	\includegraphics[width=0.32\textwidth]{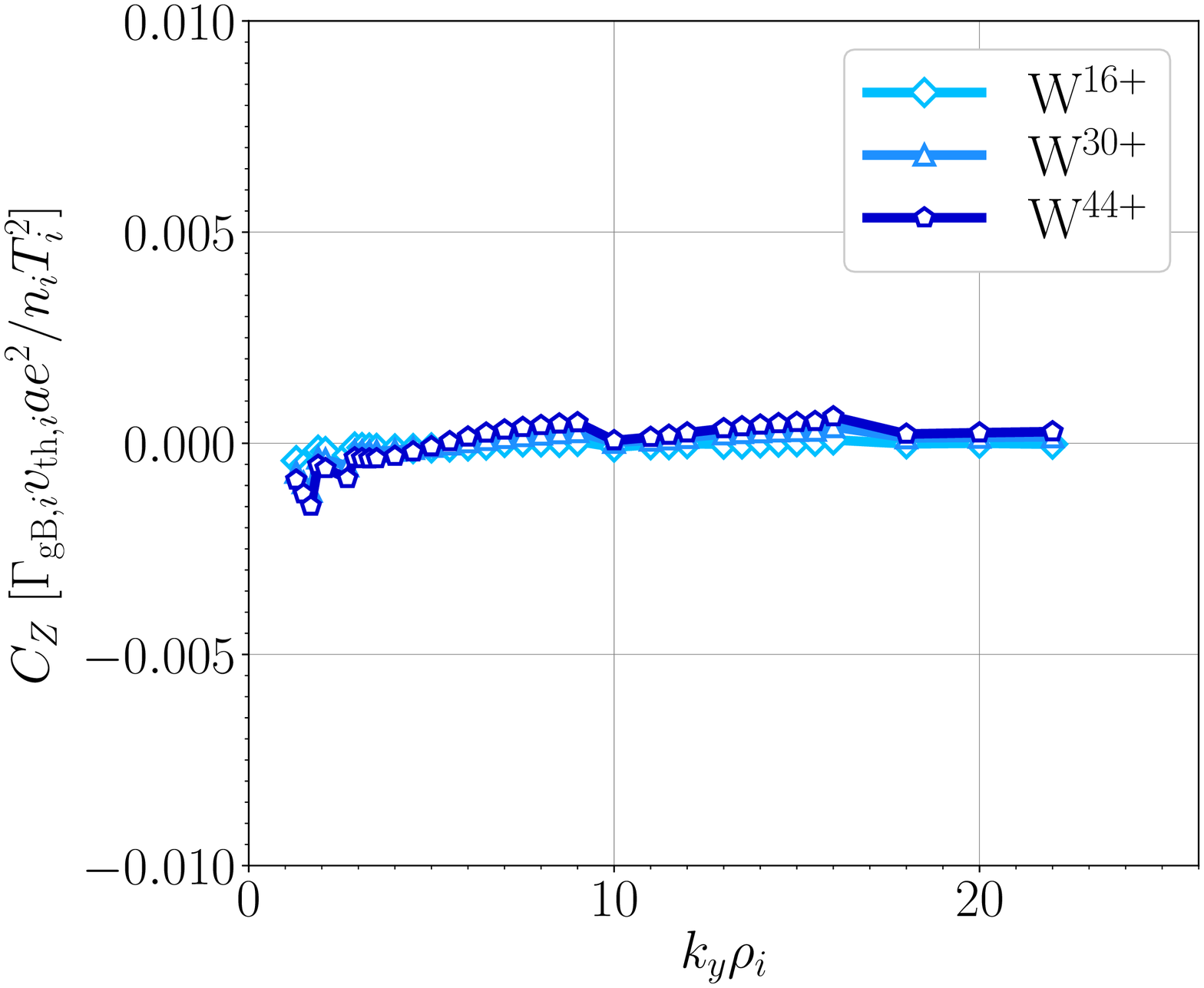}
	\caption{For the ETG case, see table \ref{table:quasilinear}, ky-spectra of the diffusion coefficient $D_{Z1}$ (left), thermo-diffusion coefficient $D_{Z2}$ (center) and pinch in the absence of gradients $C_{Z}$ (right) for the different impurities, for a mass (top row) and charge (bottom row) scan. }
	\label{fig:qletg}
\end{figure}

Finally, the results concerning the ETG instability are represented in fig.~\ref{fig:qletg}. In contrast to the ITG and TEM cases, $D_{Z1}$ is not particularly larger than $D_{Z2}$ in absolute value, and both are, in any case, considerably smaller than in the previous two cases. In addition, $C_Z$ is practically zero, which indicates that ETG driven impurity transport should be substantially smaller compared to that driven by the ITG mode or the TEM.

In summary, the quasilinear approach to the problem has delivered the following conclusions. ITG and TEM should drive most of the impurity transport by ordinary diffusion. The ITG mode seems to be prone to develop slightly peaked impurity density profiles, as \blue{$C_Z$ and $D_{Z2}$} are considerably smaller than the dominant $D_{Z1}$ and both add inward convective contributions to the total flux. The TEM case follows roughly the same characteristics, although \blue{$C_Z$ and $D_{Z2}$} are not as small compared to the diffusion coefficient $D_{Z1}$, which point out the tendency to develop peaked impurity density profiles with larger gradients than in the ITG case. In general, the sign and size of the transport coefficient are in reasonably good agreement with the analytical predictions \citep{Helander_ppcf_60_084006_2018}. The main difference resides on the mass or charge dependence of our results, which arises from the fact that all terms, including the parallel streaming neglected on the analytical treatment \green{of \citep{Helander_ppcf_60_084006_2018}}, are retained \red{in our simulations}.

\subsection{Nonlinear turbulent transport of trace impurities}
\label{sec:nonlinear}

\begin{table}
	\begin{center}
		\begin{tabular}{ccccc}
			& $a/L_{T_i}$ & $a/L_{T_e}$ & $a/L_{n_i}=a/L_{n_e}$ & $T_e/T_i$\\
			\hline
			ITG & $4.0$ & $0.0$ & $0.0$ & 1.0 \\
			TEM & $0.0$ & $0.0$ & $4.0$ & 1.0 \\
			\hline
			Species & \multicolumn{3}{c}{Ar$^{16+}$, W$^{16+}$, W$^{44+}$}\\
			\hline
		\end{tabular}
	\end{center}
	\caption{Normalized gradients, electron to ion temperature ratio, and selected impurities considered for the nonlinear transport analysis.}
	\label{table:nonlinear}
\end{table}

\begin{figure}
	\begin{center}
		\includegraphics[height=5.5cm]{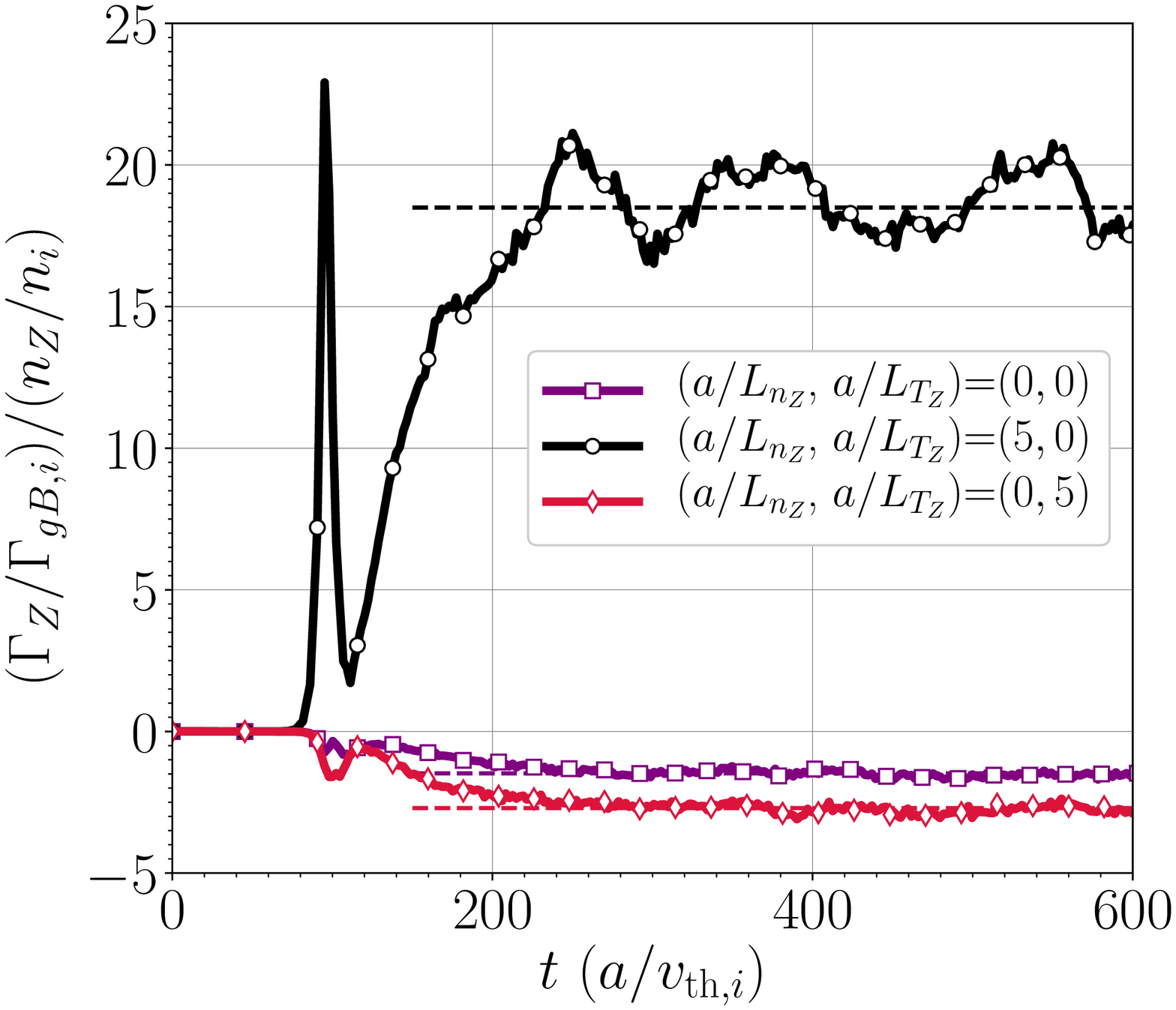}
		\includegraphics[height=5.5cm]{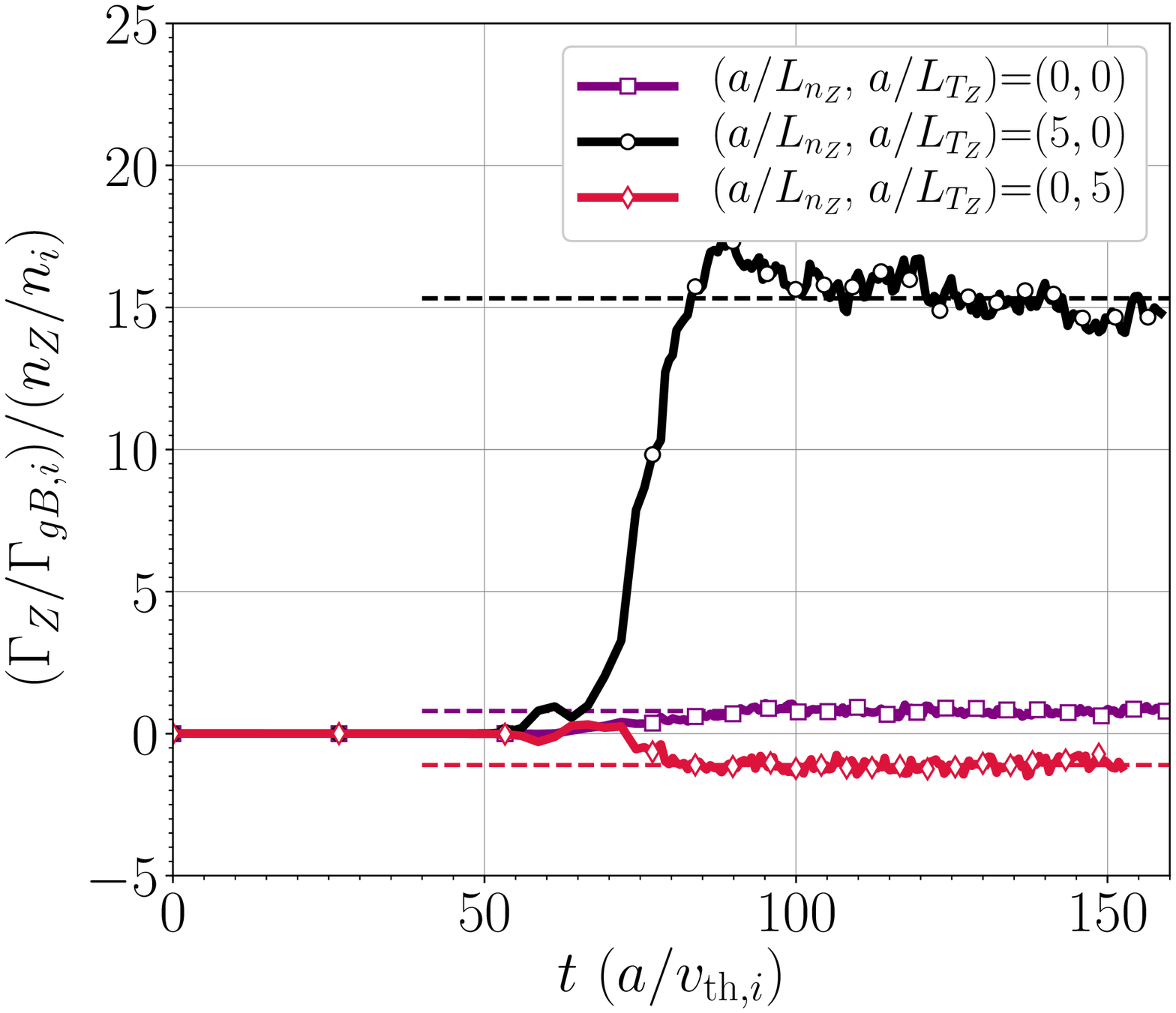}
		\caption{Normalized \blue{Ar$^{16+}$} particle flux as function of time in the presence of ITG-driven (left) and 
			TEM-\blue{driven} (right) background turbulence. The result is represented for three 
			different pairs of Ar$^{16+}$ density and temperature gradients, $(a/L_{n_Z}, a/L_{T_Z})=(0,0)$ (\blue{squares}), 
			$(a/L_{n_Z}, a/L_{T_Z})=(5,0)$ (\blue{circles}), $(a/L_{n_Z}, a/L_{T_Z})=(0,5)$ (diamonds).}
		\label{fig:3fluxes}
	\end{center}
\end{figure}

Quasilinear analyses allow \green{us} to extract qualitative information about the more or less prominent role of an instability \green{in the} turbulent impurity flux, the relative size of the different diffusive and convective terms, the direction, inward or outward, of the flux driven through each transport coefficient, and its wavenumber spectra. However, quasilinear calculations cannot provide a quantitative estimation of the flux, as nonlinear simulations do, since no saturated state is reached. \green{While nonlinear multispecies gyrokinetic simulations have been employed with remarkable success for tokamaks, see for instance \citet{Barnes_prl_2012} for a comprehensive study of the scaling of the impurity transport of particles, momentum and energy, in stellarators they are anecdotical}. In the present section, the question about the size of the turbulent impurity transport driven by ITG- and TEM-driven microturbulence and the respective transport coefficients is addressed by means of nonlinear simulations. Three of the impurities considered in the previous section, Ar$^{16+}$, W$^{16+}$ and $W^{44+}$, have been selected. Each simulation \red{includes} hydrogen nuclei, electrons and one single impurity species, all three kinetically treated. For the ITG case the resolution has been set to $N_z\times N_x\times N_y\times N_{v_{\|}}\times N_{\mu}=96\times 76\times 151\times 24\times 12$, while for the TEM turbulence $N_z\times N_x\times N_y\times N_{v_{\|}}\times N_{\mu}=96\times 76\times 256\times 48\times 12$ has been \red{taken}. The width of the box along the binormal \green{and radial directions have been set to $L_y=125\rho_i$ and $L_x=180\rho_i$, respectively}, and the flux tube has been extended one turn poloidally. \green{Standard twist-and-shift boundary conditions \citep{Beer_pop_2_2687_1995} have been considered}. In table \ref{table:nonlinear} the parameters considered for the background turbulence of interest in each case is indicated, together with the selected impurities.\\
In \red{order} to obtain the transport coefficients, $D_{Z1}$, $D_{Z2}$ and $C_Z$, for each impurity and type of background  turbulence, three simulations have been performed with different values of impurity normalized density and temperature gradients, $a/L_{n_Z}$ and $a/L_{T_Z}$, respectively. As an example, the three time traces of the turbulent flux of Ar$^{16+}$ are illustrated for the ITG case in fig.~\ref{fig:3fluxes} (left) and for TEM turbulence in fig.~\ref{fig:3fluxes} (right). For each of them, the mean value of the flux during the saturated phase is represented by a dashed line. Looking at the flux evolution when the impurity density and temperature gradients are zero (open squares linked by the purple solid line), it is immediately observed that each type of turbulence drives, \red{in the absence of impurity density and temperature gradients}, flux contributions with opposite sign. While ITG drives an inward (pinch) contribution, TEM turbulence drives outward transport (anti-pinch). Note that this first observation contradicts the quasilinear results, where in both cases $C_Z$ was positive, thus it drove negative flux for all $k_y$. However, that contribution is weak, and comparable to that arising when the temperature gradient is non-zero (open diamonds connected by the red solid line). On the other hand, the flux driven when only the density gradient is applied (open circles connected by the black solid line) is by far the largest, no matter if the background turbulence is ITG- or TEM-driven, which anticipates that ordinary diffusion will be the dominant contribution to the turbulent particle \green{flux} of impurities, as it will be quantitatively confirmed below for the other two impurity species considered.

\begin{figure}
	\begin{center}
		\includegraphics[height=5.5cm]{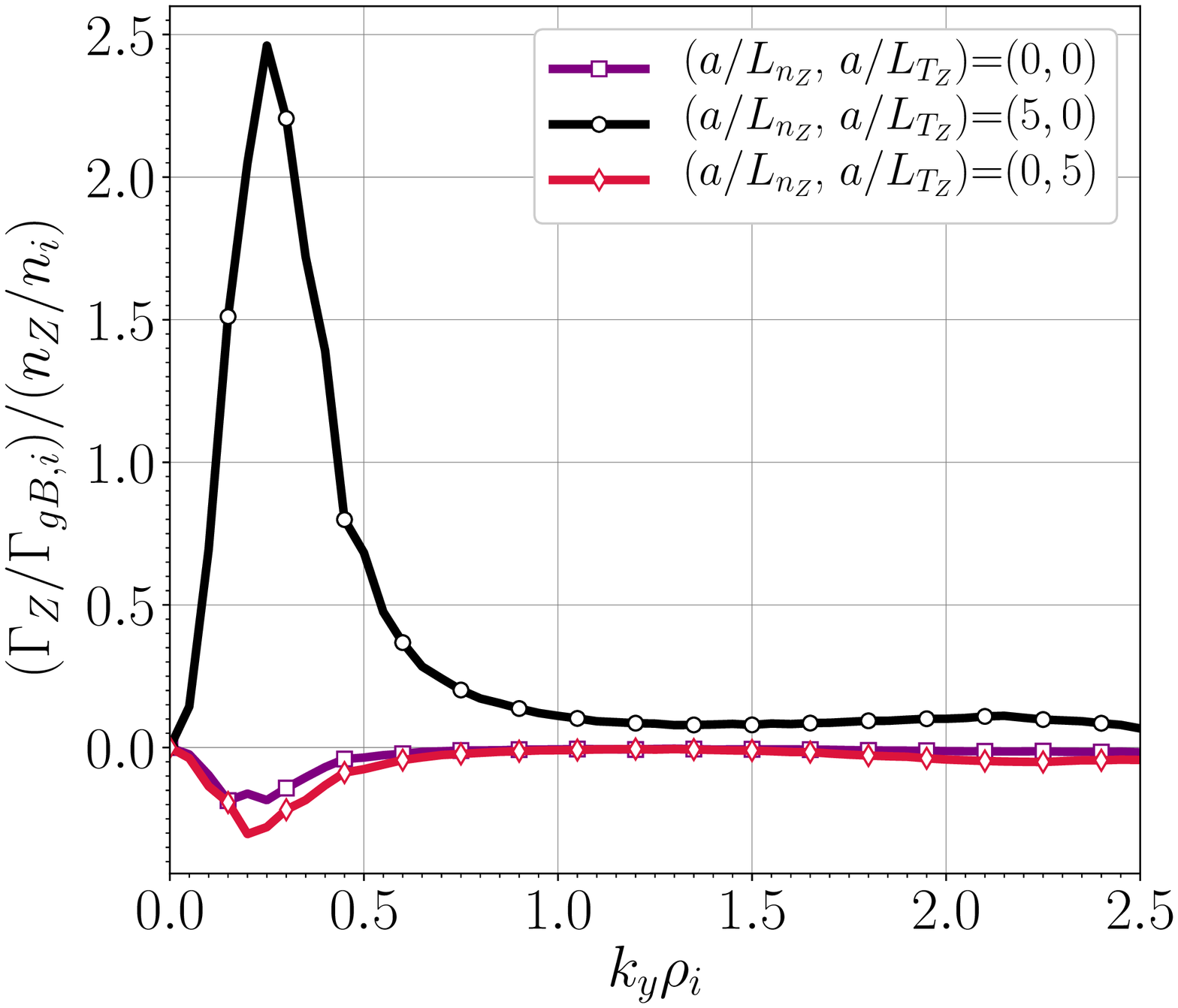}
		\includegraphics[height=5.5cm]{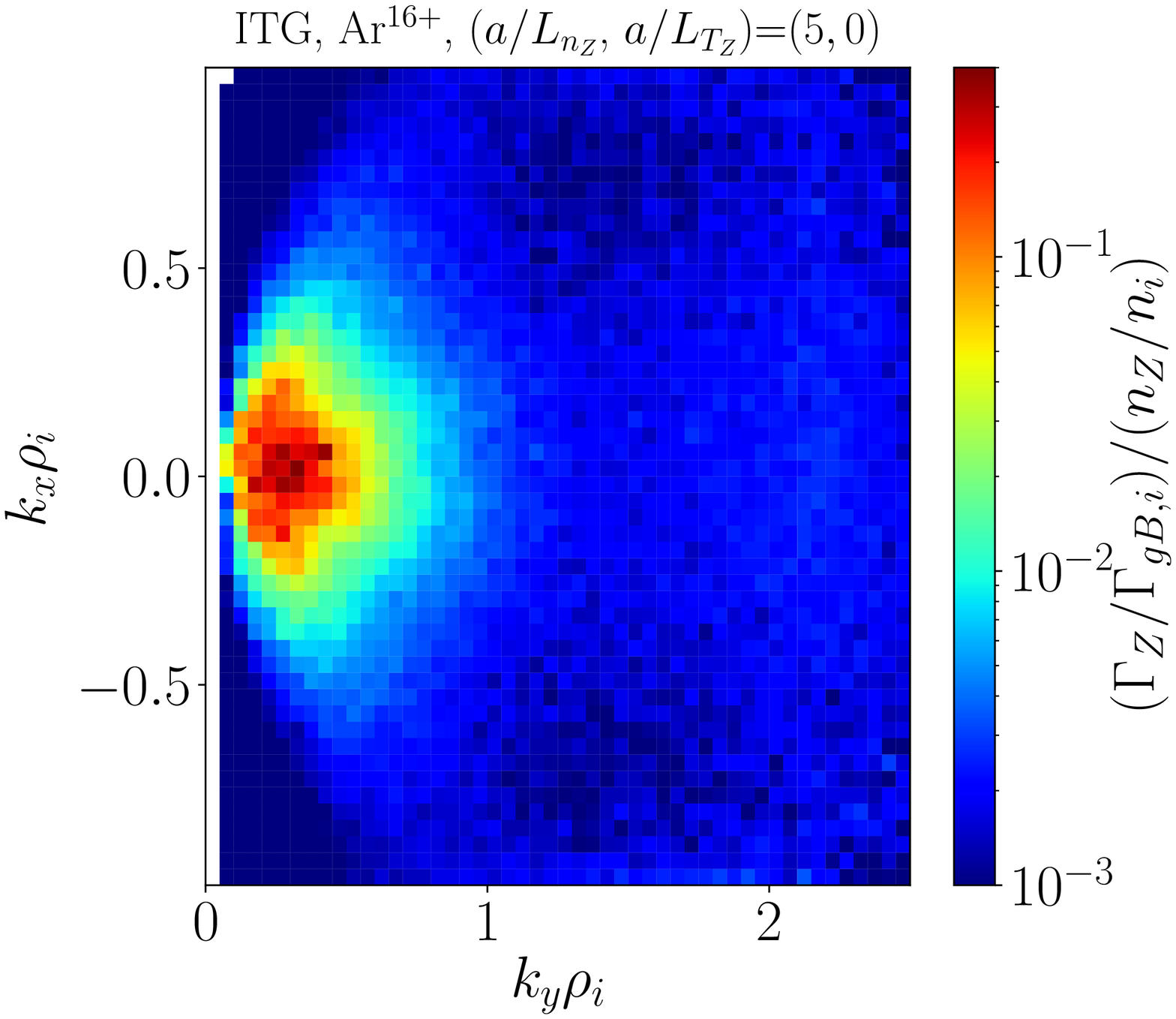}
		\caption{For ITG microturbulence: (left) $k_y$-spectrum of the normalized turbulent flux of Ar$^{16+}$ considering $(a/L_{n_Z}, a/L_{T_Z})=(0,0)$ (\blue{squares}), $(a/L_{n_Z}, a/L_{T_Z})=(5,0)$ (\blue{circles}), $(a/L_{n_Z}, a/L_{T_Z})=(0,5)$ (diamonds); (right) $(k_x,k_y)$-spectrum of the normalized turbulent flux of Ar$^{16+}$ when $(a/L_{n_Z}, a/L_{T_Z})=(5,0)$.}
		\label{fig:itgspectra}
	\end{center}
\end{figure}

Given the width of the linear growth rate spectra of the simulated ITG and TEM modes, see fig.~\ref{fig:linspectra}, one might wonder if the turbulent flux spectra are that broad or if the chosen resolution does not leave important flux contributions out of the selected range of wavenumbers. For the case of Ar$^{16+}$ embedded in ITG microturbulence the binormal wavenumber spectrum is represented for the three simulated pairs of impurity density and temperature gradients in fig.~\ref{fig:itgspectra} (left). It can be seen that most of the flux contribution arises from large scales with $k_{y}\lesssim 1$, although finite contributions can also be observed for the remaining part of the spectrum. This is particularly visible when the impurity density gradient is the only gradient applied (open circles connected by the black solid line). Regarding the radial wavenumber spectrum, fig.~\ref{fig:itgspectra} (right) shows the spectrum $k_x$ and $k_y$ of the Ar$^{16+}$ converged flux. As for the ITG-driven turbulence, the largest contributions to the flux arise from a narrow region of large \red{radial} scales, with maximum contribution along $k_x=0$. Although not shown here, these features are found with little variations for the rest of estimated fluxes under ITG conditions, independently of the impurity species.

\begin{figure}
	\begin{center}
		\includegraphics[height=5.5cm]{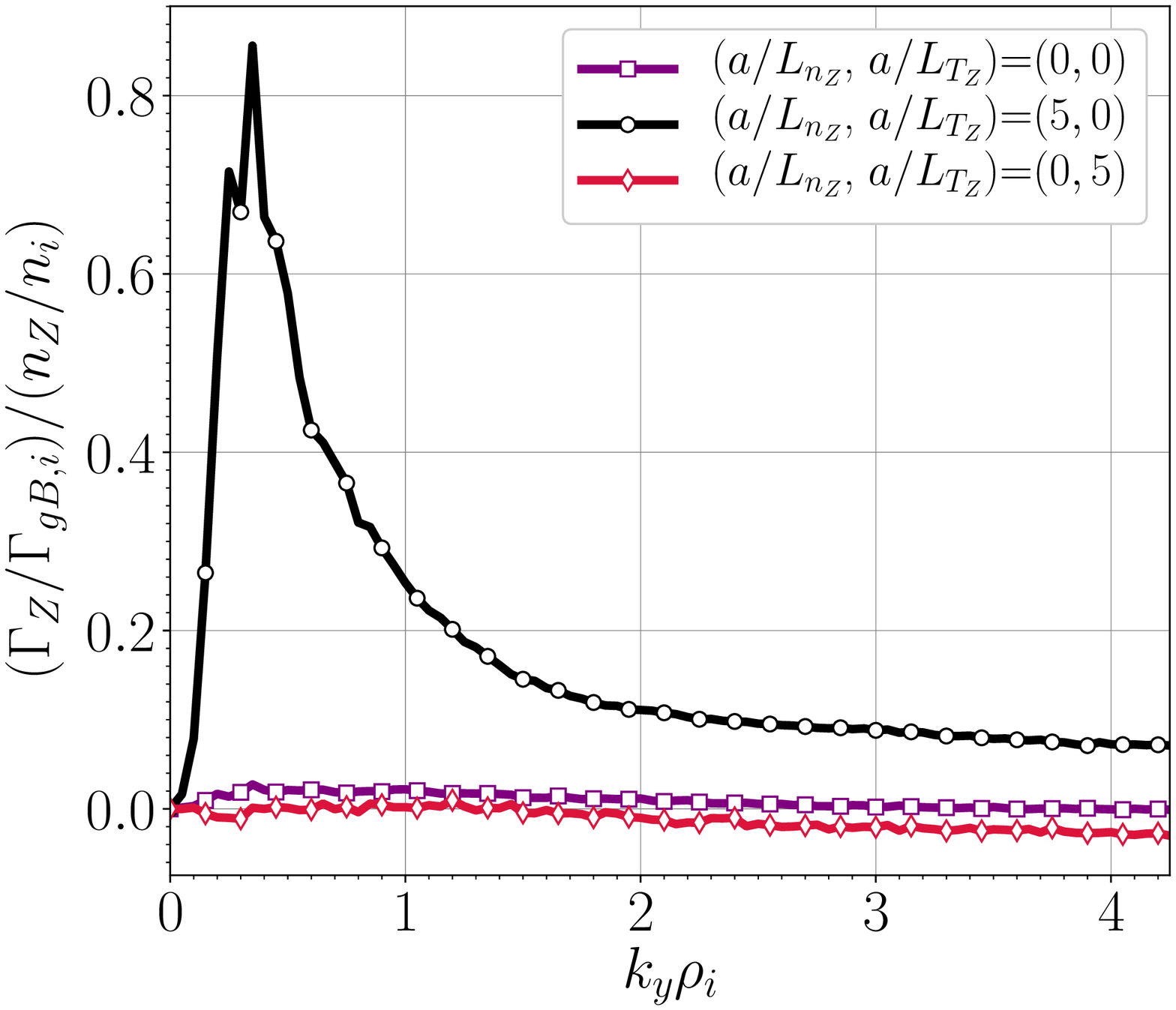}
		\includegraphics[height=5.5cm]{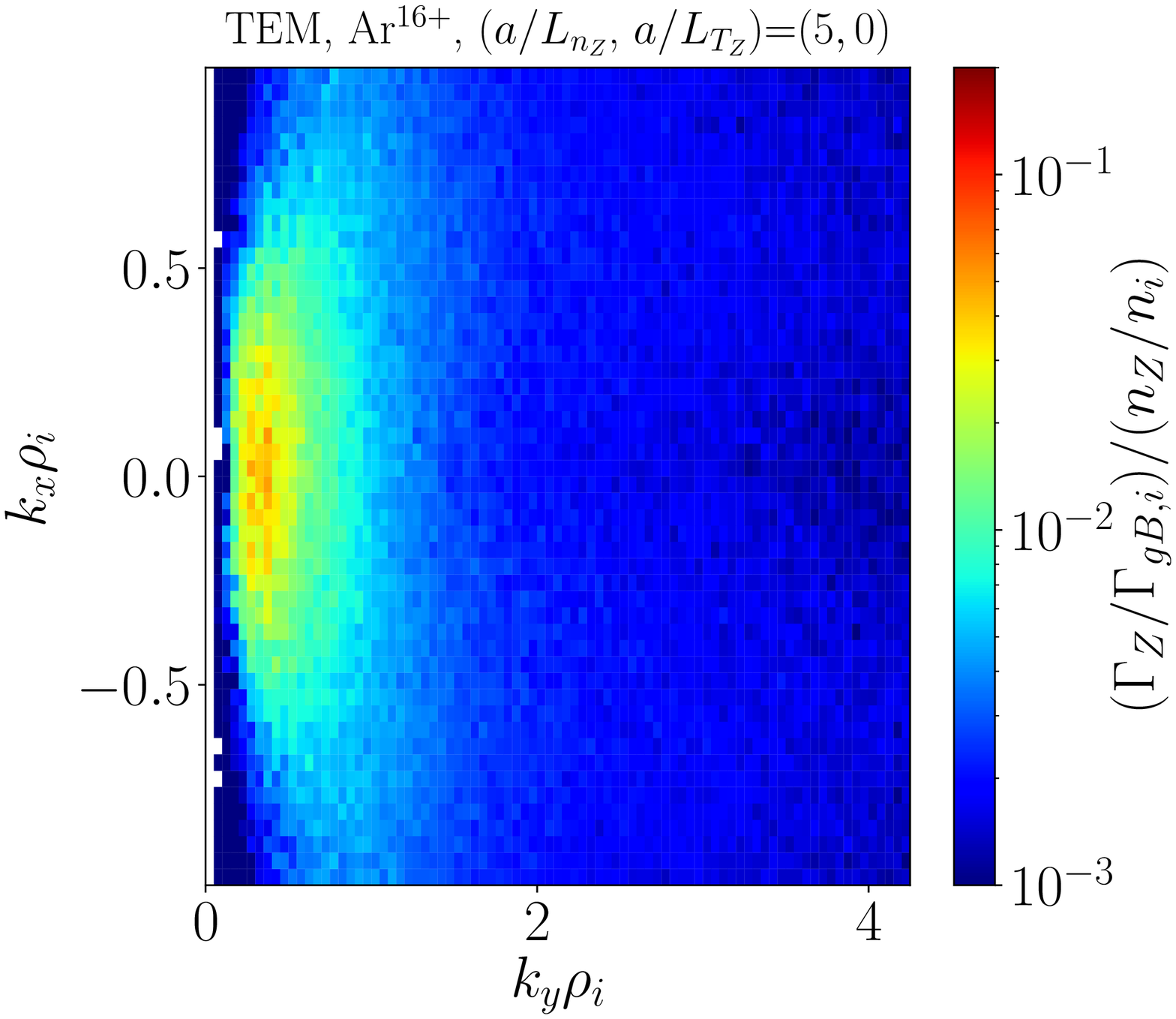}
		\caption{For TEM microturbulence: (left) $k_y$-spectrum of the turbulent particle flux of Ar$^{16+}$ considering $(a/L_{n_Z}, a/L_{T_Z})=(0,0)$ (\blue{squares}), $(a/L_{n_Z}, a/L_{T_Z})=(5,0)$ (\blue{circles}), $(a/L_{n_Z}, a/L_{T_Z})=(0,5)$ (diamonds); (right) $(k_x,k_y)$-spectrum of the turbulent particle flux of Ar$^{16+}$ when $(a/L_{n_Z}, a/L_{T_Z})=(5,0)$.}
		\label{fig:temspectra}
	\end{center}
\end{figure}

The equivalent two plots for the three simulations performed for Ar$^{16+}$ in the presence of TEM microturbulence are shown in fig.~\ref{fig:temspectra}. The subfigure on the left, representing the binormal flux spectrum, reveals qualitative differences compared to the corresponding figure of the ITG case discussed in the previous paragraph, fig.~\ref{fig:itgspectra} (left). Looking at the $k_y$-spectrum of the flux when $a/L_{n_Z}=5$, it is obvious that TEM turbulence leads to a noticeably broader flux spectrum. Although the flux of Ar$^{16+}$ finds its largest contribution at scales with \red{$k_y\rho_i\approx 1$}, the spectrum decays for increasing wavenumber less abruptly than in the ITG example. This yields appreciable flux contributions even at the largest $k_y$ represented. Concerning the radial direction, \blue{fig.~\ref{fig:temspectra}} (right) depicts, for the case $a/L_{n_Z}=5$, the flux spectrum in $k_x$ and $k_y$, \red{that exhibits a wider $k_x$ range with noticeable flux contributions} than in the ITG case as well. Finally, regarding the much weaker flux driven in the absence of impurity density and temperature gradients, although difficult to appreciate due to the much smaller amplitude, no significant contributions to the flux are present for $k_y\gtrsim 2$. In contrast, when impurity temperature gradient is set to a finite value, the flux spectrum slightly diverges as $k_y$ increases. It is worth noting that in that case two small contributions, connected to the thermo-diffusion driven pinch and the anti-pinch in absence of gradients, are opposing to each other (see discussion about fig.~\ref{fig:3fluxes} (right)), which may require a \blue{finer} resolution than the one considered.

\begin{figure}
	\begin{center}
		\includegraphics[height=5.5cm]{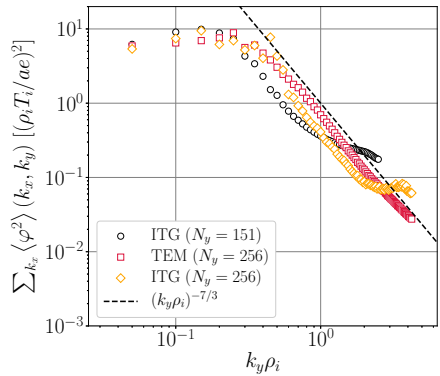}
	\end{center}
	\caption{Electrostatic fluctuation spectrum for the ITG and TEM turbulence considered for the nonlinear impurity transport calculations. Note the selected resolution choices for the transport simulations considered $N_y=151$ and $N_y=256$ for the ITG (open black circles) and TEM (open red square), respectively. A third case is considered here, \blue{an} ITG run with $N_y=256$ (open orange diamonds) for comparison. The dashed line indicated the power \blue{law} $(k_y\rho_i)^{-7/3}$.}
	\label{fig:cascade}
\end{figure}

Despite the better delimitation of the ITG-driven impurity flux spectra within the considered wavenumber window in comparison with the TEM case, it is important to emphasize that other properties of interest of the background turbulence are more solidly captured by the TEM simulations. Such is the case of the electrostatic fluctuation spectrum, which is represented, for the two \red{types} of turbulence considered throughout this section, in fig.~\ref{fig:cascade}. It can \red{be} immediately appreciated how TEM turbulent electrostatic fluctuations (open red squares) tightly follow a power law with exponent -7/3. For the ITG turbulence, the same power law is followed \red{up to} $k_y\rho_i\approx 1$, and deviates for larger values of $k_y\rho_i$. Note that this deviation is just a matter of resolution, as it shows up at lower $k_y$ values for the choice $N_y=151$ (open black circles) than for the finer resolution of $N_y=256$ (open orange diamonds). A specific investigation about why the ITG-driven particle flux spectra seem to be better bound by our mode window than for the TEM case, \red{while the energy cascade is better converged for the TEM than for the ITG case}, is out of the scope of the present paper. But, in any case, this parenthetical remark leaves us the important conclusion that both ITG and TEM microturbulence in W7-X are intrinsically three-dimensional, as demonstrated in \citet{Barnes_prl_2011}.

\begin{figure}
	\begin{center}
		\includegraphics[height=5.5cm]{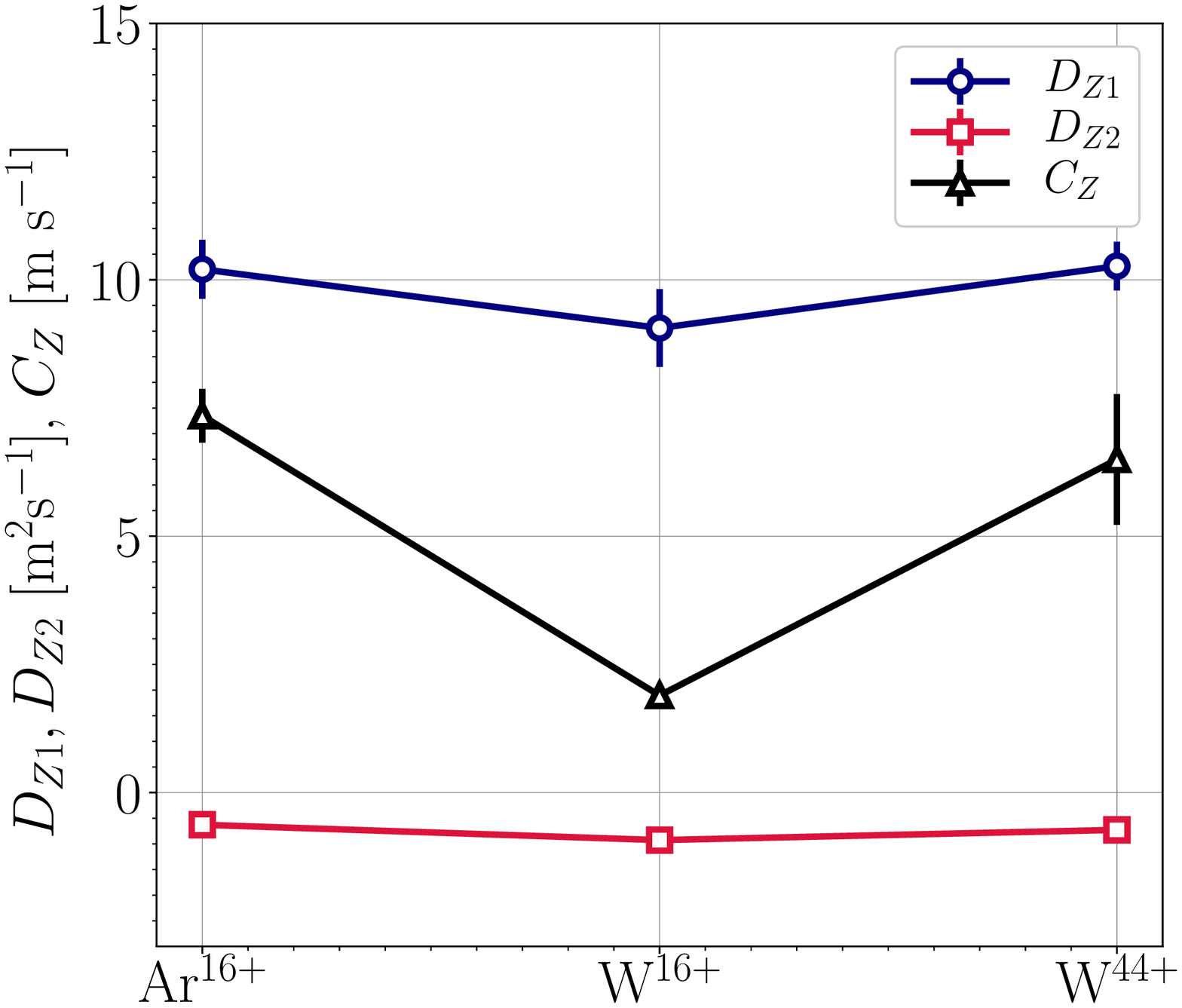}
		\includegraphics[height=5.5cm]{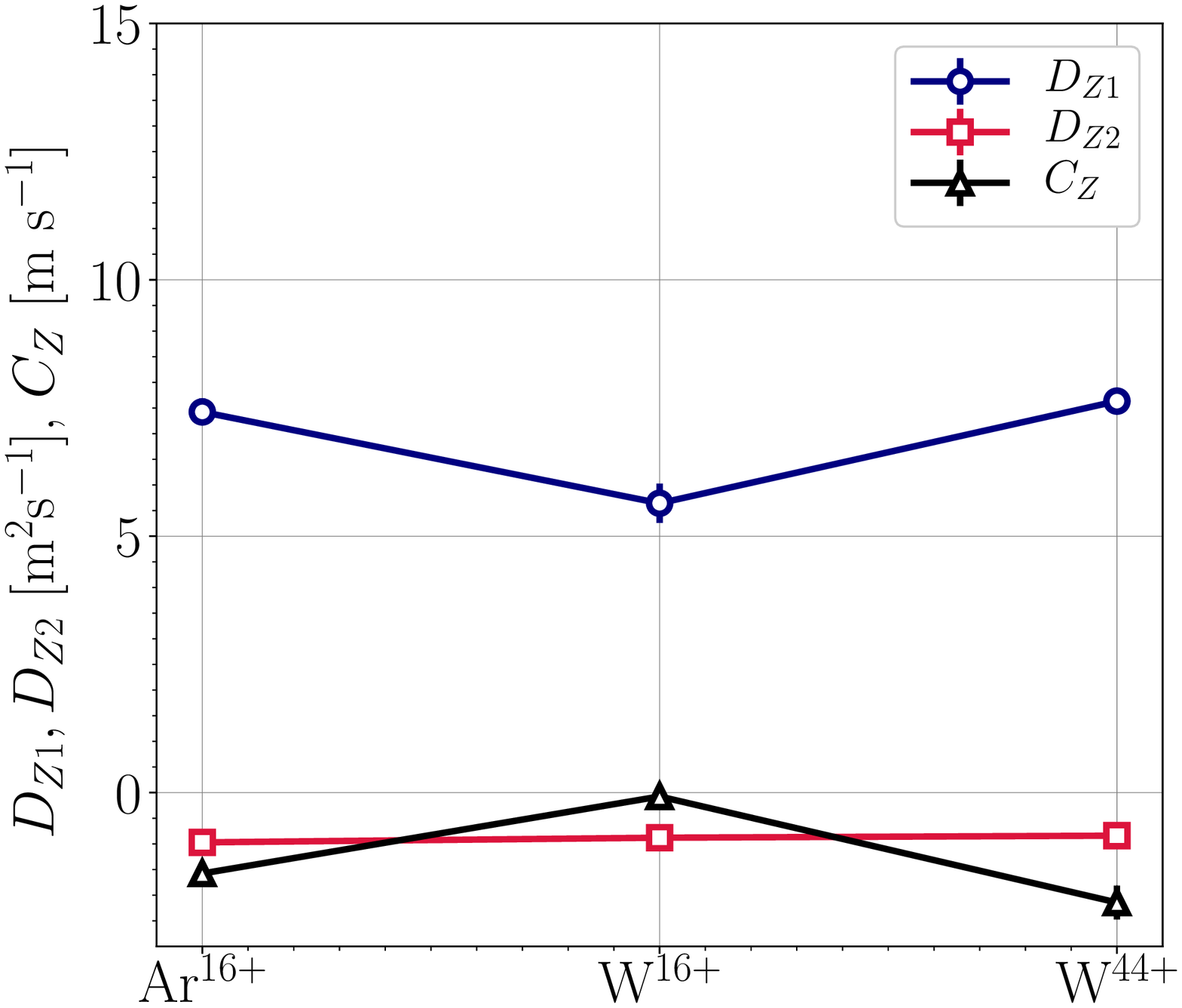}
		\caption{Diffusion coefficient, $D_{Z1}$, thermo-diffusion coefficient, $D_{Z2}$ and \blue{flux at vanishing $T'_Z$ and $n'_Z$}, $C_Z$, for Ar$^{16+}$, W$^{16+}$ and W$^{44+}$ in the presence of ITG (left) and TEM (right) microturbulence. Reference density and temperature values of $n_e=10^{19}$ m$^{-3}$ and $T_i=T_Z=1$ have been considered.}
		\label{fig:nltransport}
	\end{center}
\end{figure}

\red{Returning to the question about the size of the transport coefficients of different impurities under the influence of different type of turbulence, fig.~\ref{fig:nltransport} represents the ordinary diffusion coefficient, the thermo-diffusion coefficient and the \blue{flux in the absence of impurity and density gradient} of Ar$^{16+}$, W$^{16+}$ and W$^{44+}$} embedded in ITG microturbulence, fig.~\ref{fig:nltransport} (left), and in TEM microturbulence, fig.~\ref{fig:nltransport} (right). The reference electron density and ion temperature values considered are $n_e=10^{19}$ m$^{-3}$ and $T_i=T_Z=1$ keV, respectively. Some features common to both cases are: the dominance of the diffusion coefficient, $D_{Z1}$, above the other two coefficients, reaching values of around $10$ m$^2$s$^{-1}$ and $6-7$ m$^2$s$^{-1}$ for ITG and TEM turbulence, respectively; $D_{Z2}$ is substantially smaller than $D_{Z1}$ and adds a pinch contribution (assuming peaked $T_Z$ profiles) to the radial transport of the three species under investigation; ordinary diffusion and thermo-diffusion are practically independent on the mass and the charge state; however, the absolute value of $C_Z$ is reduced appreciably for W$^{16+}$, possibly related to its appreciably smaller charge to mass ratio compared to that for other two impurities. The only features that are clearly different for the ITG and the TEM cases are related to $C_Z$: the positive sign of $C_Z$ adds a pinch contribution in the ITG case, while the negative sign of $C_Z$ for the TEM \red{case} contributes to expulse impurities (see expression (\ref{eq:transport_law})); the absolute value of $C_Z$ is noticeably larger for ITG than for TEM. Of all these features, it is worth mentioning that the large relative size of $D_{Z1}$ or the low size and sign of $D_{Z2}$ are qualitative characteristics advanced by the quasilinear analysis. On the other hand, the relative strength between the ITG- and TEM-driven $D_{Z1}$ as well as the sign and size of $C_Z$ for the TEM turbulence are not captured by the quasilinear simulations. 

Finally, it is important to recall that the equilibrium impurity density gradient is determined by the value of the peaking factor, which is expressed as the ratio of the total convection velocity, $V$, and the diffusion coefficient $D$.  In \blue{terms} of the three coefficients under discussion, the peaking factor reads as:

\begin{equation}
\frac{V}{D}=-\frac{D_{Z2} \dd\ln T_Z/\dd r +C_Z}{D_{Z1}}.
\label{eq:peaking}
\end{equation}

In practical terms, the numerical demonstration of the large diffusion coefficient just shown yields the conclusion that microturbulence, of ITG and TEM kind, should tend to form impurity density profiles close to flatness\footnote{\blue{Note that the value of the diffusion coefficients obtained are in qualitative agreement with the experimentally measured and far above the neoclassically estimated, see \citet{Geiger_NF_59_046009_2019}, where the diffusion coefficient of LBO-injected iron impurities is found to be up to approximately $3$ m$^2$s$^{-1}$ in the radial position we are simulating. The larger values of the numerically calculated diffusion coefficient, in particular for the ITG turbulence, can be due mainly to the two following reasons: the simulations consider a pure ITG case with a value of $a/L_{T_i}=4.0$ comparable to the experimental profiles but with $a/L_{n_i}=0$, while in the experiment $a/L_{n_i}\approx 1$, which is known to play a stabilizaing role, at least linearly \citep{Alcuson_ppcf_62_2020}; the simulations are performed for a limited region in $\alpha$ (for the most unstable flux tube) which might lead to interpret the numerical value more as an upper bound than as actual estimation for a specific flux surface, over which the impurities are actually distributed in real experiments.}}. For instance, for the values shown in fig.~\ref{fig:nltransport} (left) the resulting peaking factor in equilibrium, although negative for ITG background conditions, would difficultly reach large absolute values, unless the impurity \blue{temperature} gradient were unrealistically strong. The peaking factor would be even weaker in the presence of TEM turbulence, since it exhibits a rather weak anti-pinch \blue{at vanishing $T'_Z$ and $n'_Z$ together with a} pinch contribution of comparable size driven by thermo-diffusive processes,  which would in the end lead to a peaking factor fairly close to zero. {On the other hand, the fact that $C_Z$ results in an outward contribution to the flux opens the possibility that TEM drives hollow impurity density profiles, \green{and motivates a deeper investigation of the properties of this pinch on the magnetic configuration space of W7-X \citep{Alcuson_inprogress}}.


\subsection{\red{Nonlinear turbulent transport of non-trace impurities}}
\label{sec:nonlinear2}

\begin{figure}
	\begin{center}
		\includegraphics[height=5.5cm]{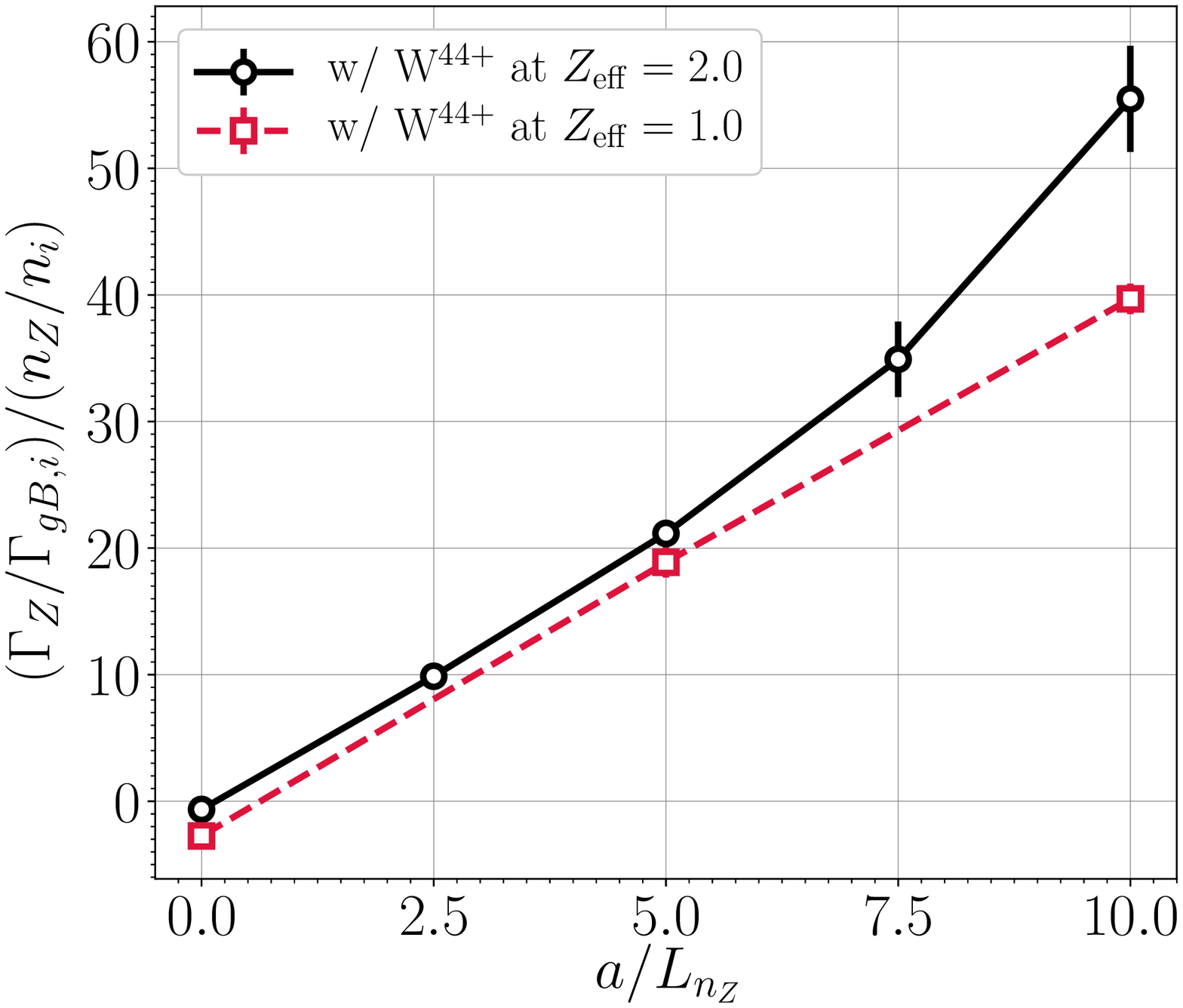}
		\includegraphics[height=5.5cm]{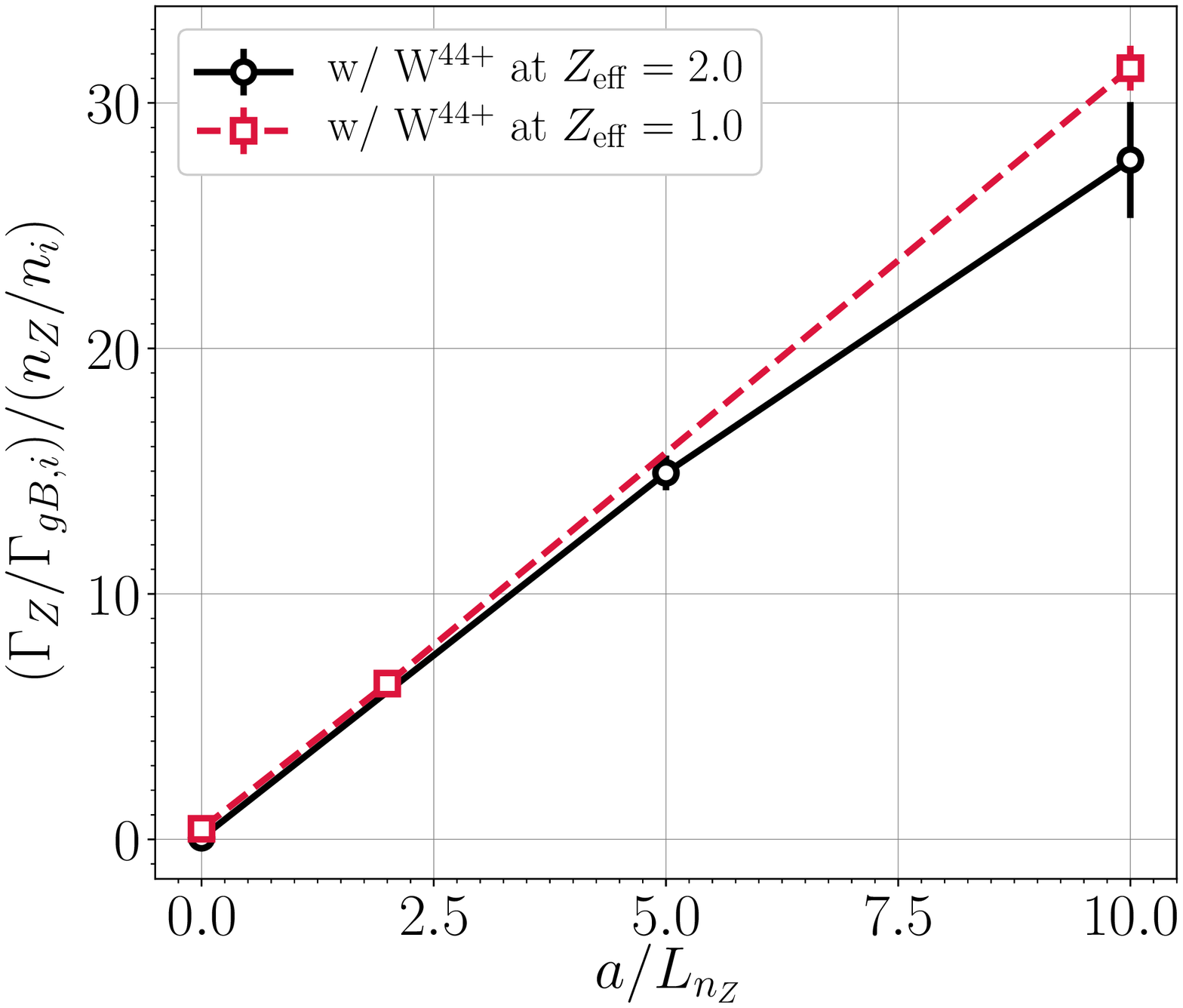}
		\caption{Normalized radial particle flux  of W$^{44+}$ as a function of its normalized density gradient at trace concentration (open squares) and at a concentration that makes $Z_{\mathrm{eff}}=2.0$ (open circles), for ITG (left) and TEM (right) microturbulence.}\label{fig:nontrace}
	\end{center}
\end{figure}

All the calculations up to this section have considered impurities at trace concentration, so that they responded to \red{the background turbulence driven by the bulk species gradients without affecting that \green{turbulence}. In that situation, the flux of the impurities scales linearly with their density and temperature gradients, the impurity transport coefficients are constant as long as the background turbulence is not altered, and they can be obtained employing expression (\ref{eq:transport_law})}. \red{Nonetheless, in laboratory plasmas impurities are frequently present at non-trace concentration levels, and the assumption of impurity turbulent fluxes scaling proportionally to the impurity density and temperature gradients does not necessarily hold}. \red{For this reason, the present subsection touches the question of how much the tendency of the impurity flux deviates from linear when the impurity concentration is no longer negligible. It is not the intention of the present section to provide a detailed study including several species and different background turbulence, as done in section \ref{sec:nonlinear}  for the nonlinear analysis for trace impurities. The purpose is rather to shed some light that indicates to what degree the conclusions drawn in section \ref{sec:nonlinear} can be extrapolated for realistic impurity content.} We have performed a series of simulations considering W$^{44+}$ at a concentration such that the effective charge is $Z_{\text{eff}}=2$. Only the impurity density gradient has been scanned, keeping $a/L_{T_{Z}}=0$, as we have seen that \green{density gradient drives} the dominant contribution to the turbulent flux of impurities. The resulting normalized turbulent fluxes of W$^{44+}$ are represented in fig.~\ref{fig:nontrace} (left) for the ITG-driven background turbulence and in fig.~\ref{fig:nontrace} (right) for the TEM \red{case}. For the curves representing the flux of W$^{44+}$  at $Z_{\text{eff}}=2$, it can be observed that the deviation from the linear trend is only noticeable  at rather large normalized density gradient values, larger than $a/L_{n_Z}\approx 5$. This deviation is more obvious for the ITG case that for the TEM, and each of them points to opposite effects: while the TEM-driven turbulent transport of \red{tungsten} tends to be weakened \red{with} respect to the linear behaviour, the ITG-driven flux is enhanced. Apart from that, the presence of non-trace \red{tungsten} introduces an offset respect to the linear trend in the ITG case, that is not found for the TEM. In other words, the presence of \red{tungsten} at non-trace concentration is altering the value of the ITG-driven pinch \blue{in the absence of tungsten density and temperature gradients} towards making it nearly zero, as can be noted looking at the two points represented for $a/L_{n_Z}=0$. In any case, a closing remark from these simulations is that, unless $Z_{\text{eff}}$ is much larger than $2$, \red{the dependence of the impurity fluxes on the impurity density gradient seems close enough to linear so that the conclusions drawn in section~\ref{sec:nonlinear} can be extrapolated to moderately realistic values of $Z_{\text{eff}}$}.

\section{Conclusions}
\label{sec:conclusions}
In the present work, the transport of impurities driven by gyrokinetic microturbulence has been \red{investigated} for W7-X geometry. Quasilinear calculations and nonlinear simulations have been performed in the flux tube and electrostatic approximations with the recently developed code \stella. The transport coefficients of several trace impurities in the presence of ITG, TEM and ETG unstable conditions have been analyzed. The ETG, only considered in the quasilinear analysis, has shown substantially smaller impurity transport coefficients compared to the ITG and TEM cases. The conclusions drawn from the nonlinear results for ITG and TEM microturbulence \red{indicate} that, independently on the charge and the mass of the impurity, the turbulent transport is dominated by ordinary diffusion, and that thermo-diffusion contributes very weakly to push the impurities radially inward. The estimated diffusion coefficient has been found to be in qualitative agreement with the experimentally reported for W7-X plasmas. The contribution driven in absence of gradients, $C_Z$, has been found to be a pinch in the presence of ITG microturbulence and an anti-pinch under the influence of the TEM conditions. These features, some of them qualitatively anticipated by the quasilinear calculations, translate into an optimistic picture of the transport of impurities in W7-X, where the large microturbulence driven diffusion would contribute to produce nearly flat equilibrium impurity density profiles, free of strong radial localization of impurities. The possible extrapolation of these conclusions to realistic non-trace concentration of impurities has been partially confirmed by simulations at $Z_{\text{eff}}=2$, that have demonstrated that the diffusion coefficient does not deviates substantially from a linear dependence on the impurity density gradient.

\section*{Acknowledgements}
\label{sec:acknowledgements}
This work has been carried out within the framework of the EUROfusion Consortium and has received funding from the Euratom research and training programme 2014-2018 and 2019-2020 under grant agreement No. 633053. The views and opinions expressed herein do not necessarily reflect those of the European Commission. This research was supported in part by grant PGC2018-095307-B-I00, Ministerio de Ciencia, Innovaci\'on y Universidades, Spain. The simulations were carried out in the clusters Marconi (Cineca, Italy) and Xula (Ciemat, Spain). J. M. Garc\'ia-Rega\~na is grateful to A. Ba\~n\'on-Navarro for helpful discussions.
\section*{References}

$\bibliographystyle{jpp}

\nocite{*} 

\end{document}